\documentclass[10pt]{article}
\usepackage{graphicx}

\def\Title#1{\begin{center} {\Large #1 } \end{center}}
\def\Author#1{\begin{center}{ \sc #1} \end{center}}
\def\Address#1{\begin{center}{ \it #1} \end{center}}

\newcommand\pubblock{\rightline{\begin{tabular}{l} Proceedings of the Fifth Annual LHCP\\ \pubnumber\\
         \pubdate  \end{tabular}}}

\newenvironment{Abstract}{\begin{quotation} \begin{center} 
             \large ABSTRACT \end{center}\bigskip 
      \begin{center}\begin{large}}{\end{large}\end{center} \end{quotation}}

\newenvironment{Presented}{\begin{quotation} \begin{center} 
             PRESENTED AT\end{center}\bigskip 
      \begin{center}\begin{large}}{\end{large}\end{center} \end{quotation}}

\def\beq{\begin{equation}}
\def\eeq#1{\label{#1}\end{equation}}
\def\eeqn{\end{equation}}

\def\beqa{\begin{eqnarray}}
\def\eeqa#1{\label{#1}\end{eqnarray}}
\def\eeqan{\end{eqnarray}}

\let\bar=\overbar

\def\Dslash{\not{\hbox{\kern-4pt $D$}}}
\def\dslash{\not{\hbox{\kern-2pt $\del$}}}

\def\msb{{\bar{\ssstyle M \kern -1pt S}}}

\usepackage{color}
\def\deriv {\ensuremath{\mathrm{d}}}

\newcommand\pubnumber{ LHCb-PROC-2017-037 }

\newcommand\pubdate{October 14, 2017}

\def\affiliation{
On behalf of the LHCb Experiment, \\
School of Physical Sciences\\
University of Chinese Academy of Sciences, Beijing, 100049, China}

\usepackage{lineno}
\begin{document}
\large
\begin{titlepage}
\pubblock

\vfill
\Title{  New results on heavy-flavour in heavy-ion collisions with LHCb  }
\vfill

\Author{ Jiajia Qin  }
\Address{\affiliation}
\vfill
\begin{Abstract}

Heavy-flavour quarks are important to probe Quark-Gluon Plasma(QGP) properties. Cold Nuclear Matter(CNM) effects can be accessed by $p$Pb collisions. 
LHCb is a heavy-flavour precision experiment and has collected large collision data samples.
Production cross-section measurements of prompt $D^{0}$ at $\sqrt{s_\mathrm{NN}}$= 5 TeV and $J/\psi$ at $\sqrt{s_\mathrm{NN}}$= 8.16 TeV are presented. 

\end{Abstract}
\vfill

\begin{Presented}
The Fifth Annual Conference\\
 on Large Hadron Collider Physics \\
Shanghai Jiao Tong University, Shanghai, China\\ 
May 15-20, 2017
\end{Presented}
\vfill
\end{titlepage}
\def\thefootnote{\fnsymbol{footnote}}
\setcounter{footnote}{0}
%

\normalsize 


\section{Introduction}

In nucleus-nucleus collisions, a deconfined thermodynamic system, the so-called Quark-Gluon Plasma(QGP), is expected to be formed. 
Heavy-flavour quarks are sensitive to measure the properties of the medium since they are
produced at an early stage of the collision and, consequently, interact with the created deconfined system. 
To fully understand the information on the thermodynamic system provided by heavy-quark observables, effects that are not related to the creation of a deconfined system
need be well studied.
They can be studied with $p$Pb collisions.

Production cross-section measurements of prompt $D^{0}$ at $\sqrt{s_\mathrm{NN}}$= 5 TeV~\cite{5tevD} and $J/\psi$~\cite{8.16tevjpsi} at $\sqrt{s_\mathrm{NN}}$= 8.16 TeV in $p$Pb collisions
are reported in these proceedings. 
The experimental data is confronted with different types of phenomenological calculations.
The nuclear modification of the parton distribution function(PDF)~\cite{nPDFs} 
are extracted based on nuclear collision data under the assumption of collinear factorisation. They show modifications of the parton distribution function compared to the one of the free proton and neutron.
At low longitudinal momentum
fractions x carried by the parton,
collinear factorisation may break down and the effective field theory, the so-called Color Glass Condensate(CGC)~\cite{CGC}, may provide the more appropriate description of the nucleus as a dense gluonic system.
In the Coherent energy loss~\cite{loss} model, it was proposed that the dominant nuclear modification in quarkonium production might be small angle gluon radiation of the incident quark pair taking properly into account the interference of gluon radiation prior and after the hard-scattering.

\section{LHCb detector}
LHCb detector~\cite{lhcb, lhcb1} is a single-armed detector covering the
pseudorapidity range $2.0<\eta<5.0$.
Due to different beam energies per nucleon for the proton and ${}^{208}\rm Pb$ beams, the LHCb detector covers
two different acceptance regions in the nucleon-nucleon rest frame:
the $1.5<y^{*}<4.5$ corresponding to the ``forward'' region and $-5.0<y^{*}<-2.5$ corresponding to the ``backward'' region. 
The rapidity $y^{*}$ is defined with respect to the direction of the proton beam.

  
\section{Prompt $D^{0}$ production in proton-lead collisions at 5 TeV}
The decay $D^{0}\to K^{\pm} \pi^{\mp}$ is used to reconstruct the signals in the proton-lead data samples collected in 2013.
The data sample corresponds to an
integrated luminosity of $1.06\pm0.02$ $\rm nb^{-1}$ for the forward region and $0.52\pm0.01$ $\rm nb^{-1}$ for the backward region.

\subsection{The single-differential cross-sections}

\begin{figure}[!htb]
\centering
\begin{minipage}[t]{0.49\textwidth}
\centering
\includegraphics[height=2in]{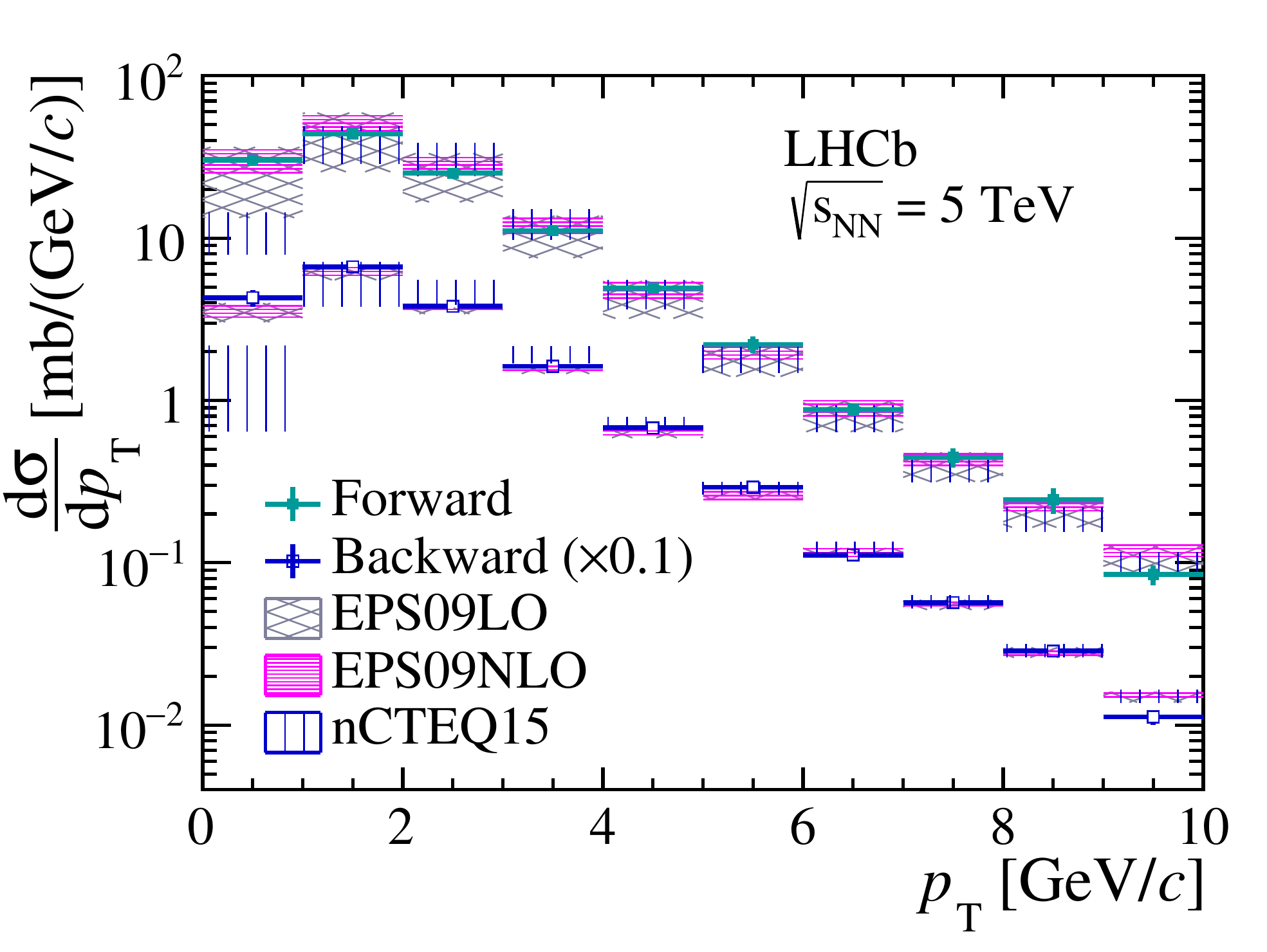}
\end{minipage}
\begin{minipage}[t]{0.49\textwidth}
\centering
\includegraphics[height=2in]{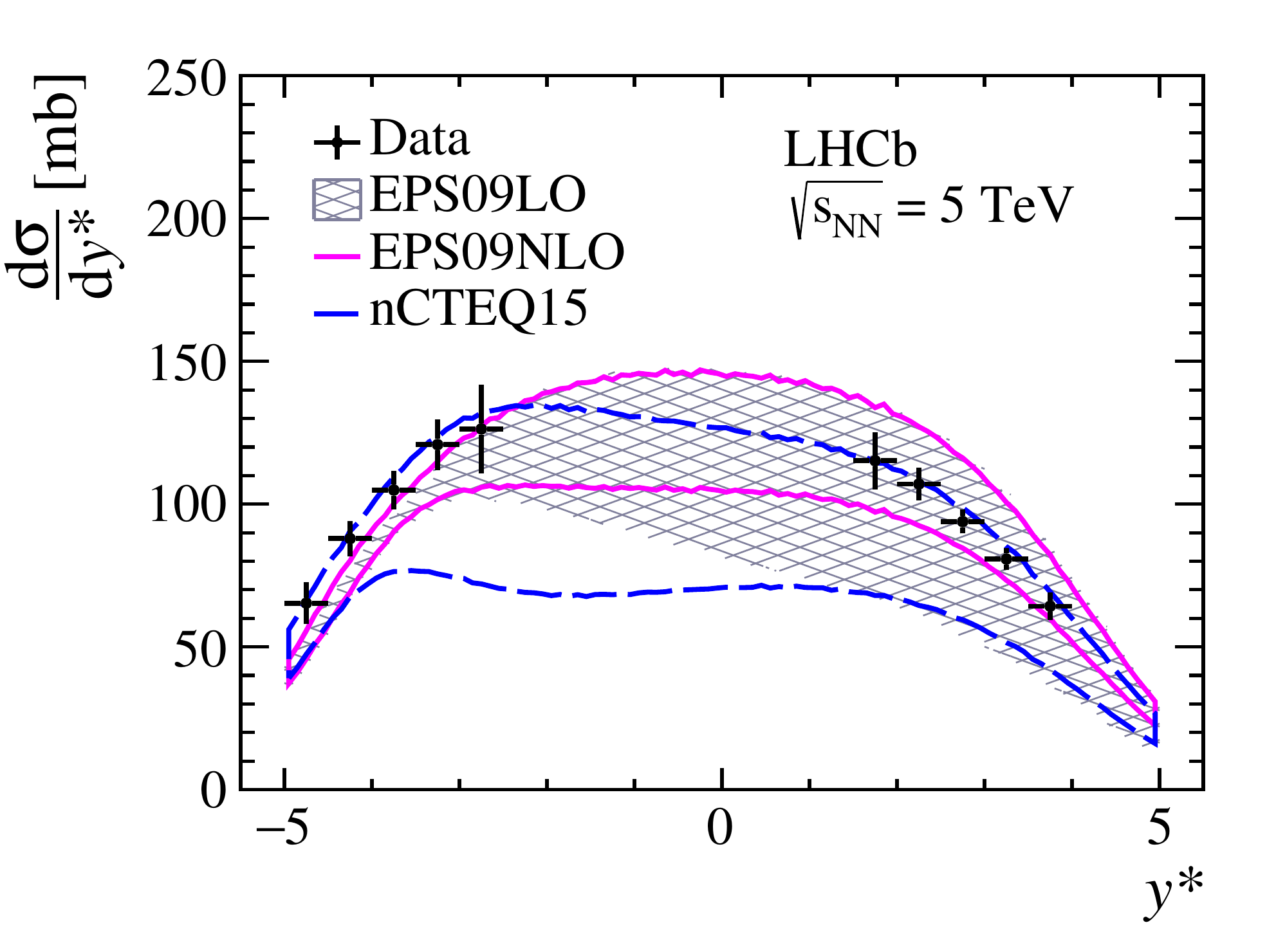}
\end{minipage}
\caption{Single-differential cross-section of prompt $D^{0}$ meson production in $p$Pb collisions (left)as a function of $p_{\rm T}$ and (right)as a function of $y^{*}$ both in the forward and backward collision samples.}
\label{figure11}
\end{figure}

The single-differential cross-sections shown in Fig.~\ref{figure11}, are compared with 
calculations(HELAC-Onia)~\cite{Shao1,Shao2,Shao3} validated with results of heavy-flavour
production cross-section
in $pp$ collisions.
The results show that the predictions based on EPS09 LO~\cite{eps09}, EPS09 NLO~\cite{eps09}, nCTEQ15~\cite{cteq} 
can give descriptions consistent
with data within large theoretical uncertainties. While the nCTEQ15 underestimates data in the low $p_{\rm T}$ region.
The authors of nCTEQ15 pointed out themselves~\cite{cteq} that the gluon distributions are hardly constrained by any data and that, hence, the nPDF uncertainties are albeit their already large size still systematically underestimated.

\subsection{Nuclear modification factors}
\begin{figure}[!htb]
\centering
\begin{minipage}[t]{0.3\textwidth}
\centering
\includegraphics[height=1.5in]{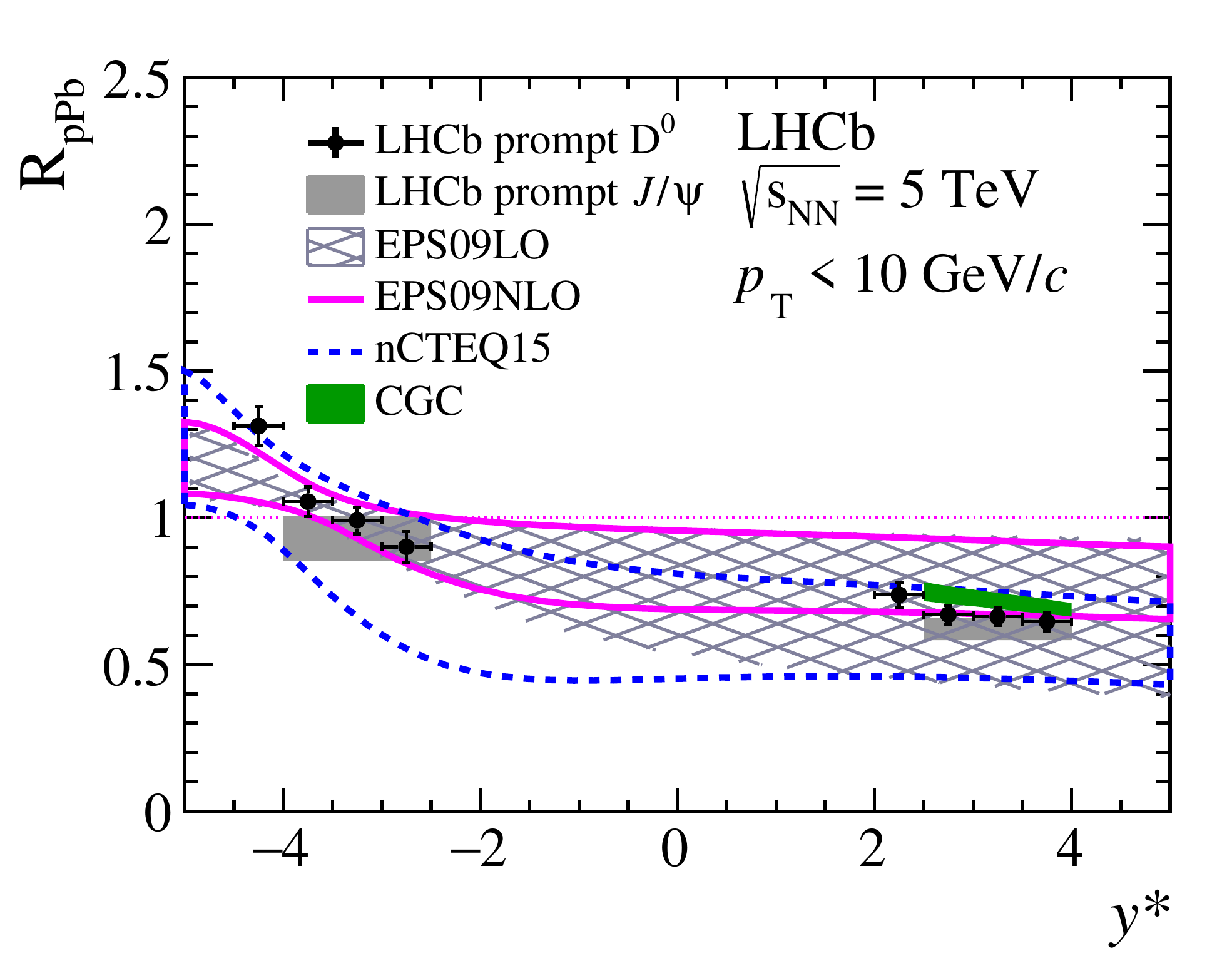}
\end{minipage}
\begin{minipage}[t]{0.3\textwidth}
\centering
\includegraphics[height=1.5in]{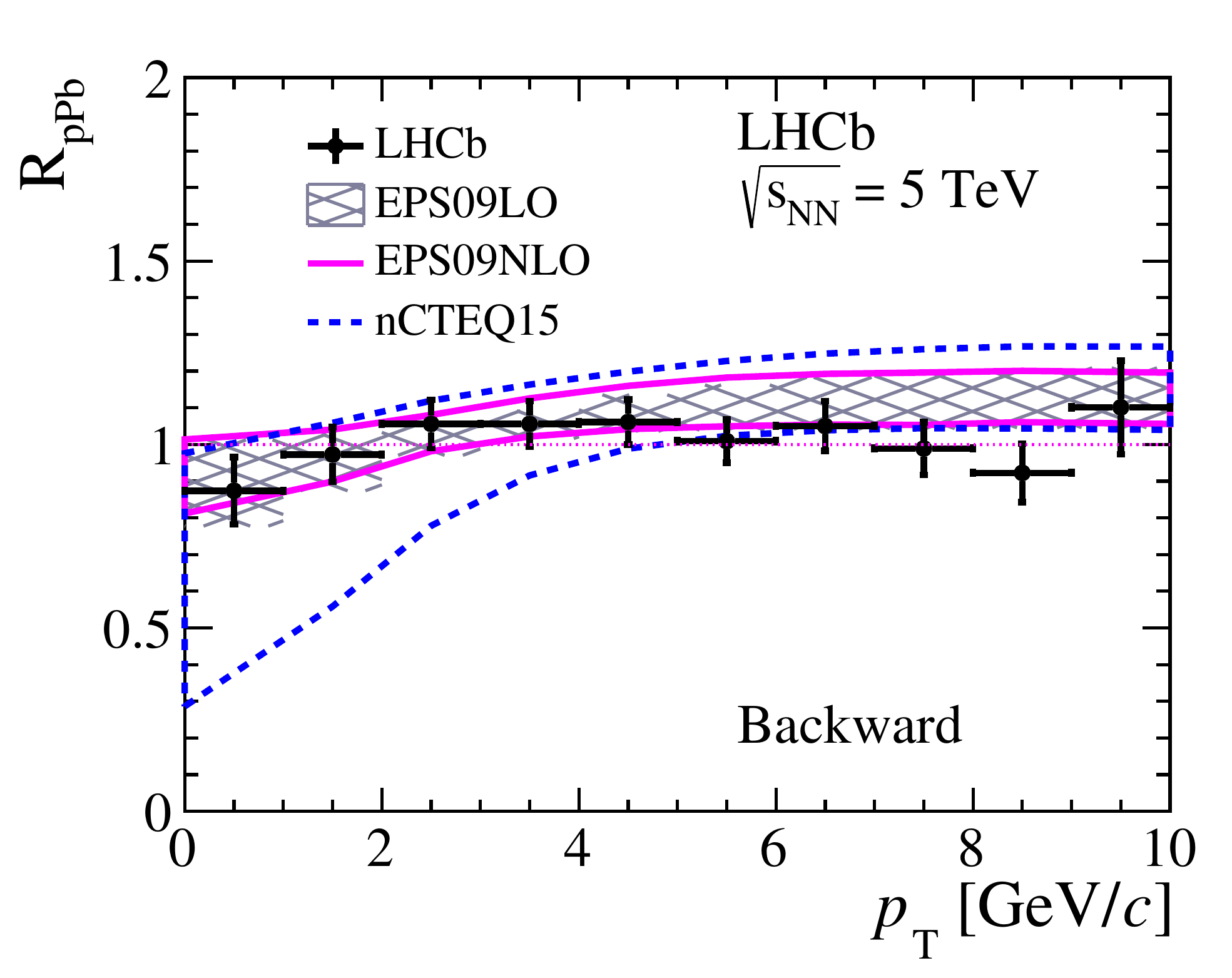}
\end{minipage}
\begin{minipage}[t]{0.3\textwidth}
\centering
\includegraphics[height=1.5in]{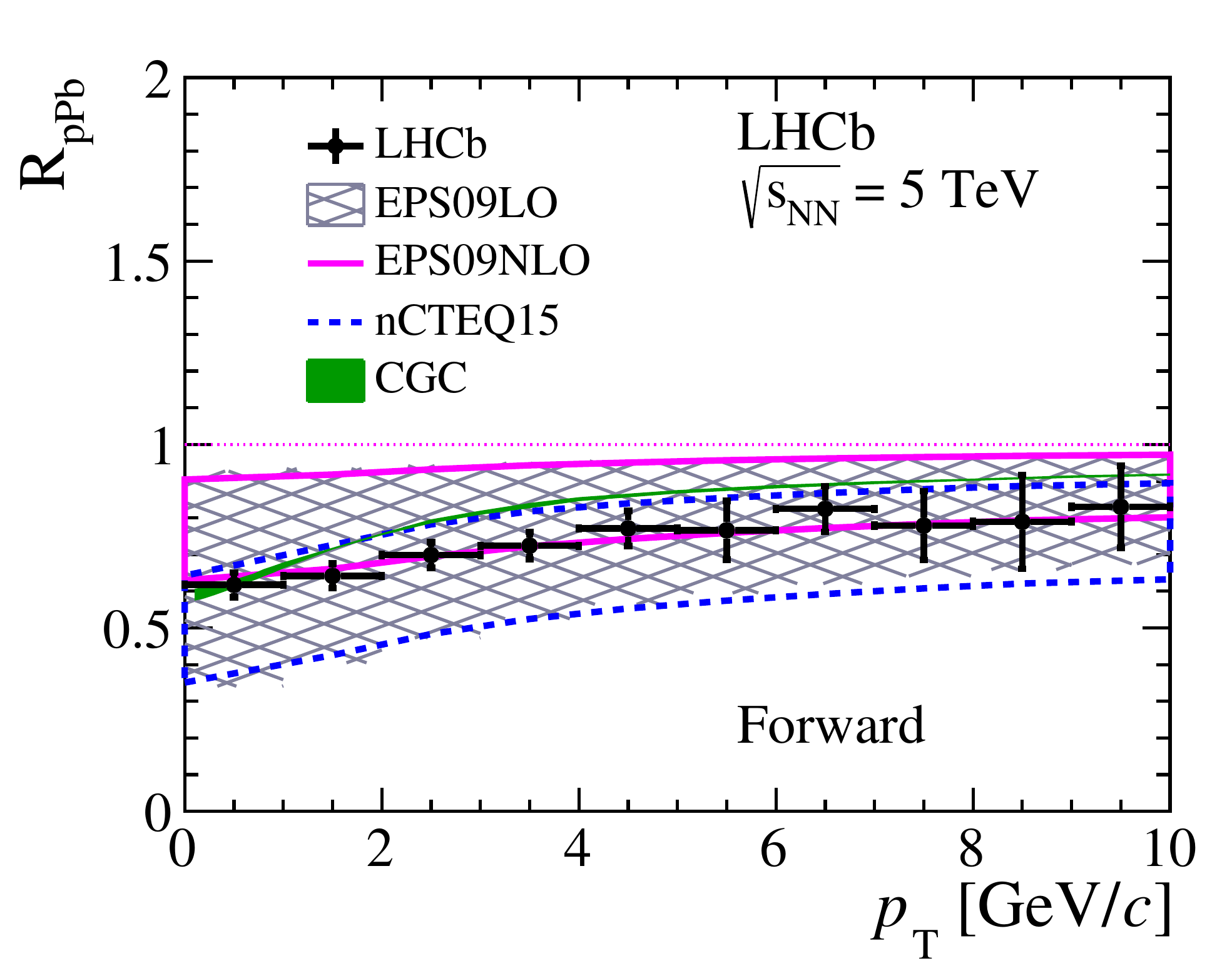}
\end{minipage}
\caption{ Nuclear modification factor $R_{p\rm Pb}$ for prompt $D^{0}$ meson production as a function of 
$y^{*}$(left) integrated up to $p_{\rm T}  = 10$ GeV/c and as a function of $p_{\rm T}$(middle and right)
integrated over the common rapidity range 
$2.5 < |y^{*}| < 4.0$ for $p_{\rm T}  < 6$ GeV/c  and 
over $2.5 < |y^{*}| < 3.5$ for $6 < p_{\rm T}  < 10$ GeV/c.}
\label{figure2}
\end{figure}

The nuclear modification factor is defined as:
\begin{equation}
R_{p\rm Pb}(y^{*}(p_{\rm T}),\sqrt{s_\mathrm{NN}}) = 
\frac{1}{A} \frac{\frac{\deriv \sigma_{pPb}}{\deriv
    y^{*}(\deriv p_{\rm T})}(y^{*}(p_{\rm T}),\sqrt{s_\mathrm{NN}})}{\frac{\deriv \sigma_{pp}}{\deriv y^{*}(\deriv p_{\rm T})}(y^{*}(p_{\rm T}),\sqrt{s_\mathrm{NN}})},
\end{equation}
where A=208 is the atomic mass number of the lead nucleus.
The results integrated over $y^{*}$ or integrated over $p_{\rm T}$ both in the forward region and the backward region are shown in Fig.~\ref{figure2}.
The results are compared with HELAC-Onia calculations using EPS09LO, EPSNLO and nCTEQ15 nPDFs~\cite{Shao1,Shao2,Shao3}.
Calculations based on CGC model~\cite{jpsicgc} are also shown in Fig.~\ref{figure2}.

For $R_{p\rm Pb}$ as a function of $y^{*}$, there is a strong suppression in the forward region,
a nuclear modification factor above unity
in the backward region.
For the results in the backward region, all three
predictions show reasonable agreement with each other and with LHCb data.
In the forward region, the HELAC calculations using nCTEQ15 and EPS09LO nPDFs show a
better agreement with the data than the calculation with EPS09NLO.
The CGC model is found to be able to describe the
trend of prompt $D^{0}$ meson nuclear modifications as a function of $p_{\rm T}$ and of rapidity.
The nuclear modification factors for prompt $D^{0}$ are also compared with those for
prompt $J/\psi$ in the left plot of Fig.~\ref{figure2} as a function of rapidity integrated over $p_{\rm T}$, and they are found to be consistent.
  
\subsection{Forward-Backward production ratio}
\begin{figure}[!htb]
\centering
\begin{minipage}[t]{0.49\textwidth}
\centering
\includegraphics[height=2in]{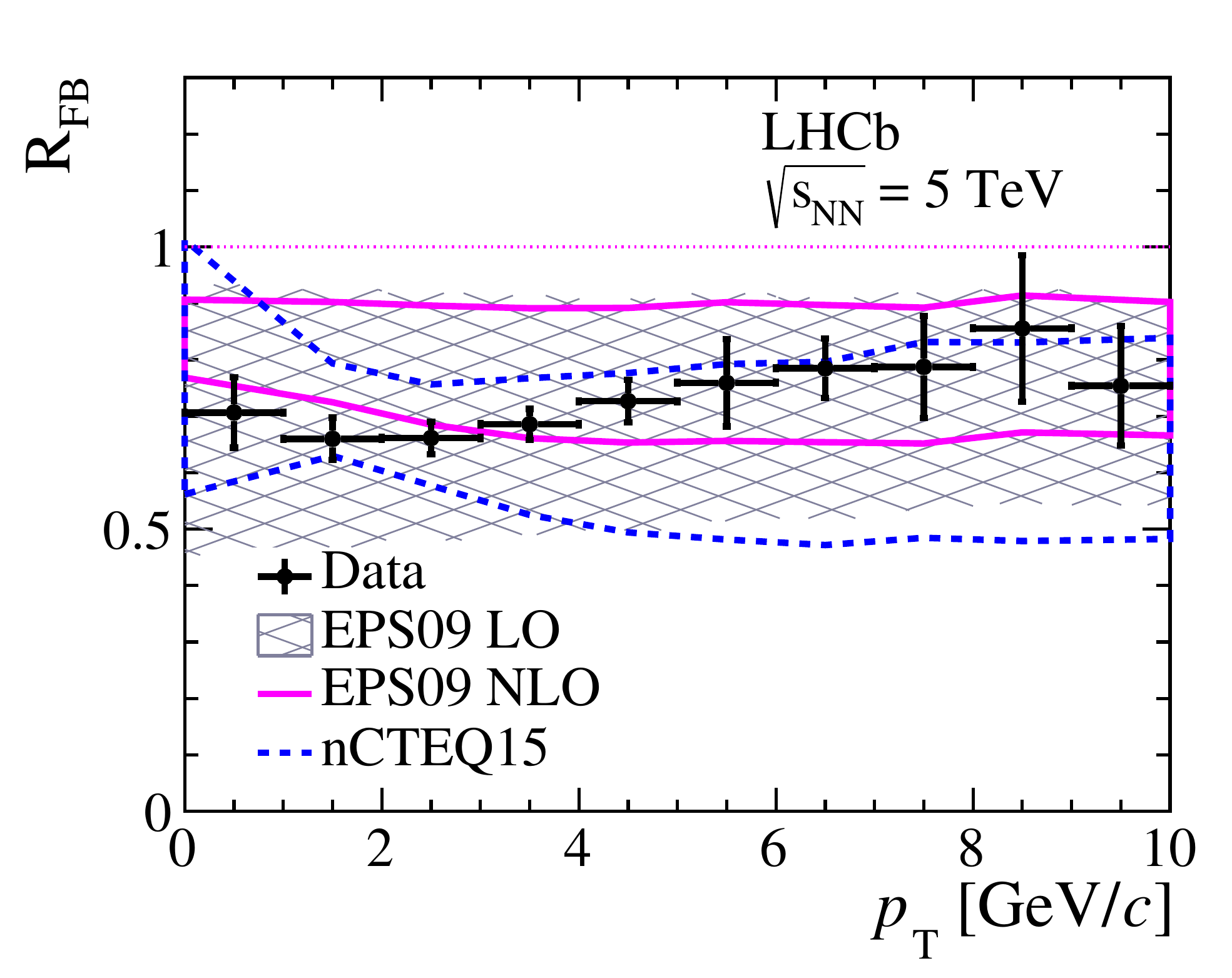}
\end{minipage}
\begin{minipage}[t]{0.49\textwidth}
\centering
\includegraphics[height=2in]{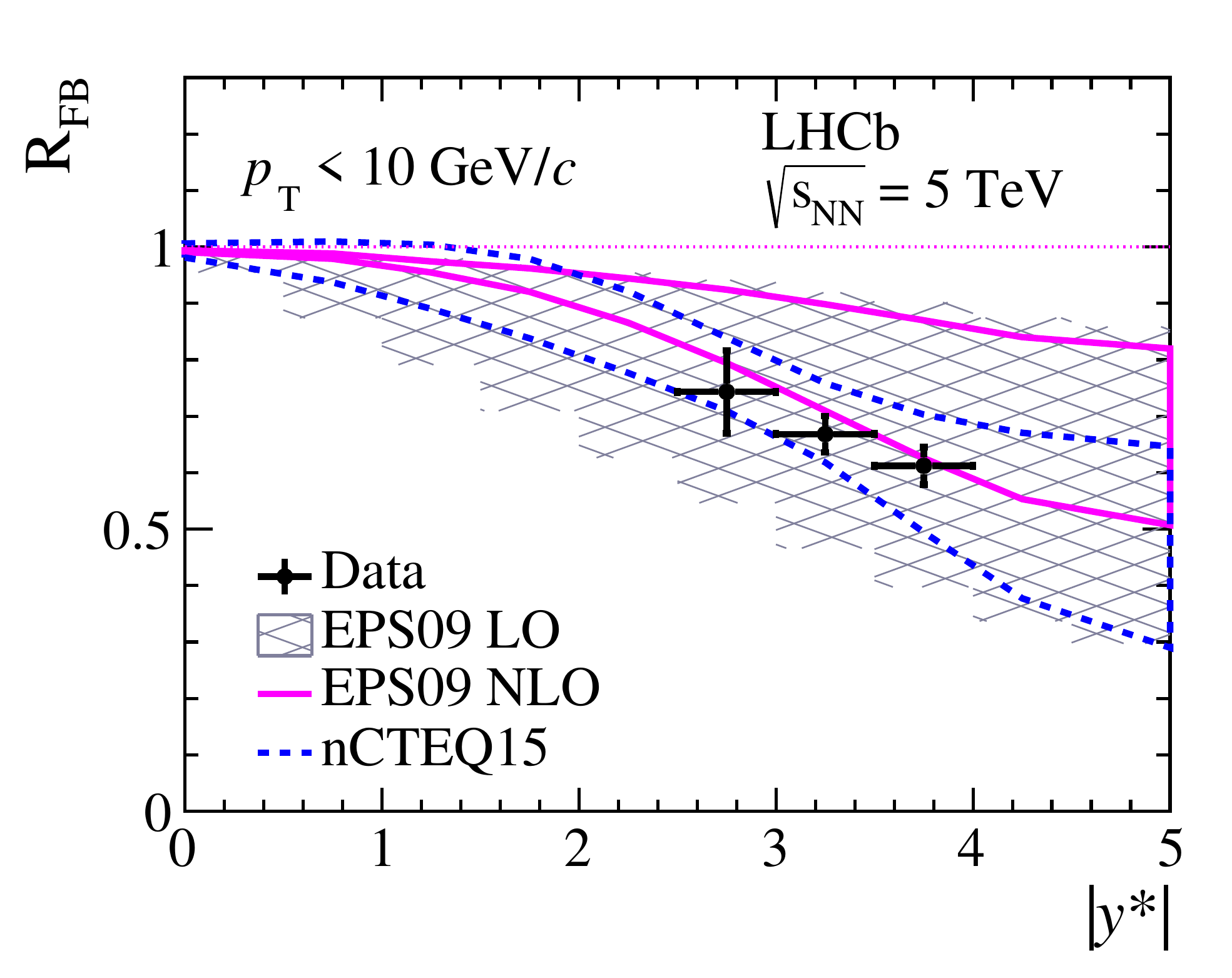}
\end{minipage}
\caption{Forward-backward ratio $R_{\rm FB}$ for prompt $D^{0}$  meson  production as a function of 
$p_{\rm T}$(left) integrated over the common rapidity range $2.5 <|y^{*}|< 4.0$ for $p_{\rm T}<6$ GeV/c and over
$2.5 <|y^{*}|< 3.5$ for $6 < p_{\rm T} < 10$ GeV/c; as a function of $y^{*}$(right) integrated up to $p_{\rm T}$ = 10 GeV/c.}
\label{figure3}
\end{figure}

\begin{figure}[!htb]
\centering
\begin{minipage}[t]{0.49\textwidth}
\centering
\includegraphics[height=2in]{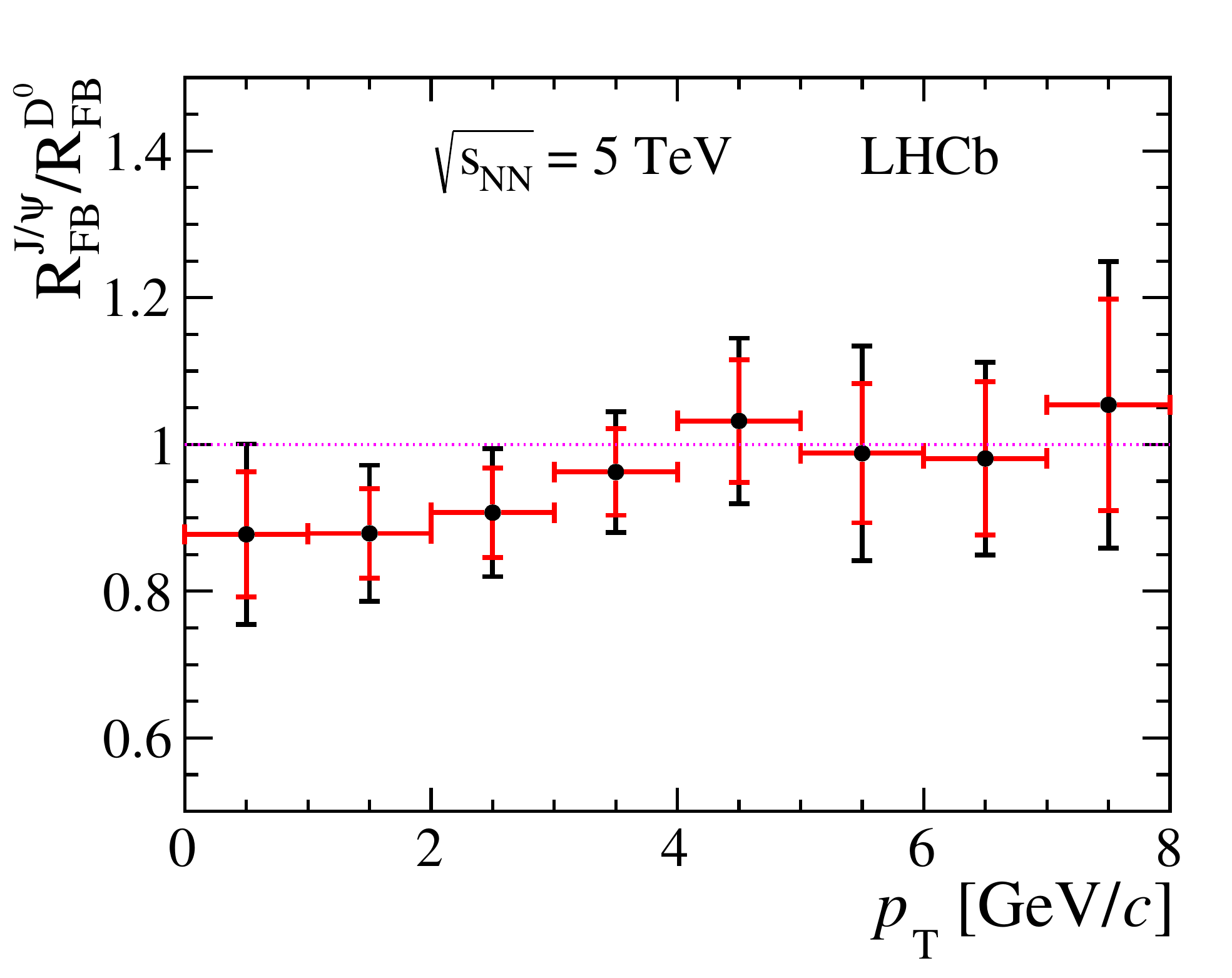}
\end{minipage}
\begin{minipage}[t]{0.49\textwidth}
\centering
\includegraphics[height=2in]{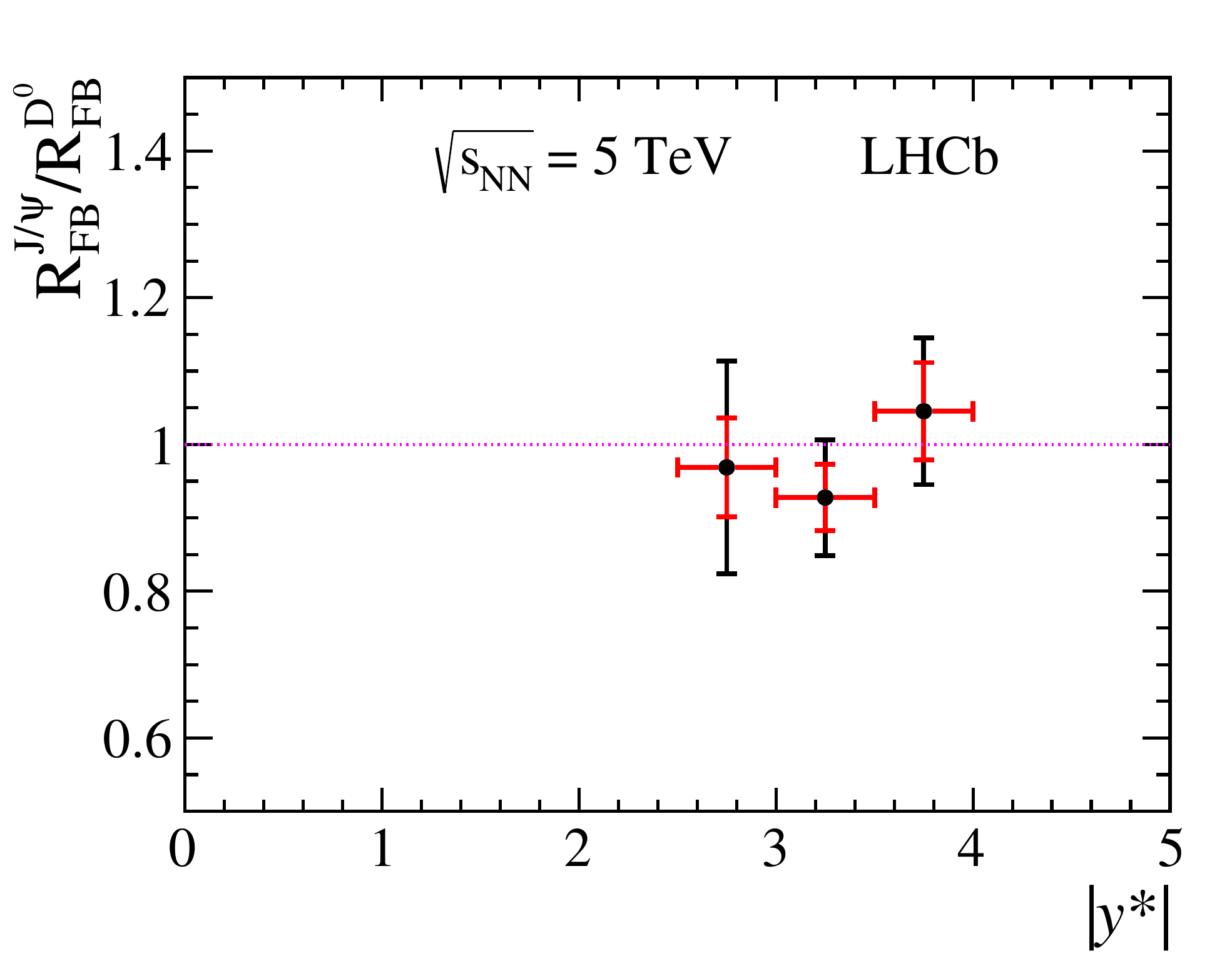}
\end{minipage}
\caption{The ratio of the forward-backward ratio with respect to prompt $J/\psi$ as a function of $p_{\rm T}$(left) and as a function of $y^{*}$(right).}
\label{figure31}
\end{figure}

The forward-backward production ratio $R_{\rm FB}$ is defined as:
\begin{equation}
R_\mathrm{FB}(y^{*}(p_{\rm T})) 
\equiv R_{p\rm Pb}(+|y^{*}|(p_{\rm T})) / R_{p\rm Pb}(-|y^{*}|(p_{\rm T})).
\end{equation}
In this quantity, a large fraction of the systematic uncertainties cancel.
The results are shown in Fig.~\ref{figure3}.
The ratio of $R_{\rm FB}$
for prompt $J/\psi$ divided by $R_{\rm FB}$ for prompt $D^{0}$ mesons is shown in Fig.~\ref{figure31}.

Good agreement is found between measurements and HELAC-Onia
predictions using EPS09LO~\cite{eps09} and nCTEQ15~\cite{cteq} nPDFs.
The forward-backward production ratio increases
slightly with increasing $p_{\rm T}$
and decreases strongly with increasing rapidity $|y^{*}|$.
The measurement shows that $R_{\rm FB}$ has the same size for prompt $D^{0}$ and prompt $J/\psi$ mesons
within the uncertainties in the LHCb kinematic range.

\section{$J/\psi$ production in proton-lead collisions at 8.16 TeV}
The production of $J/\psi$ meson was studied in $p$Pb collisions at the centre-of-mass
energy per nucleon pair $\sqrt{s_\mathrm{NN}}$ = 8.16 TeV.
Prompt $J/\psi$ and nonprompt $J/\psi$ are separated with the pseudo-proper decay time $t_{z}$.
This measurements are based on the proton-lead collision data samples collected in 2016.
The integrated luminosity corresponds to $13.6\pm0.3$ \rm $\rm nb^{-1}$ for the forward region and $20.8\pm 0.5$ \rm $\rm nb^{-1}$ for the backward region.

\subsection{Single-differential cross-sections}

\begin{figure}[!htb]
\centering
\begin{minipage}[t]{0.49\textwidth}
\centering
\includegraphics[height=2in]{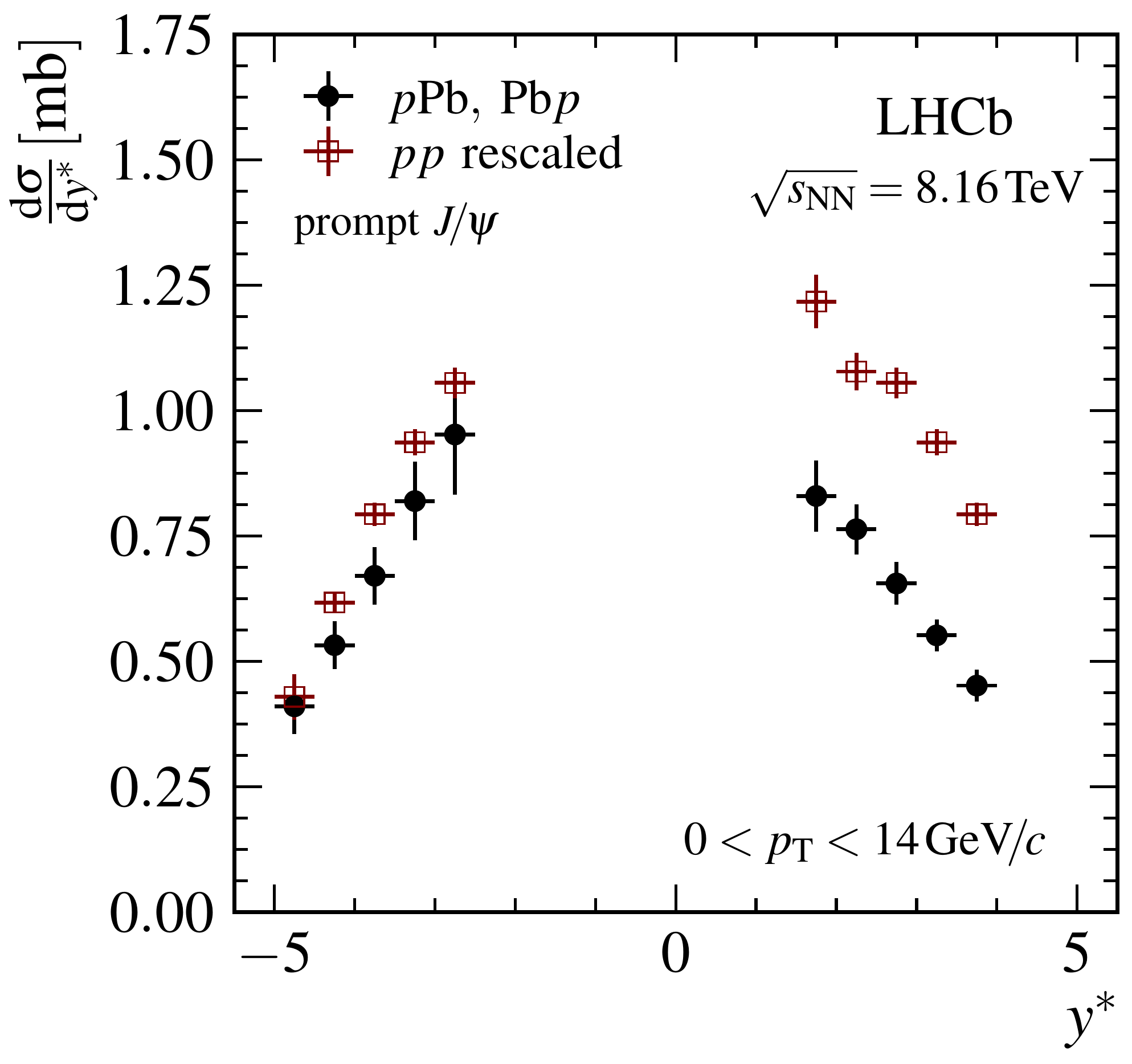}
\end{minipage}
\begin{minipage}[t]{0.49\textwidth}
\centering
\includegraphics[height=2in]{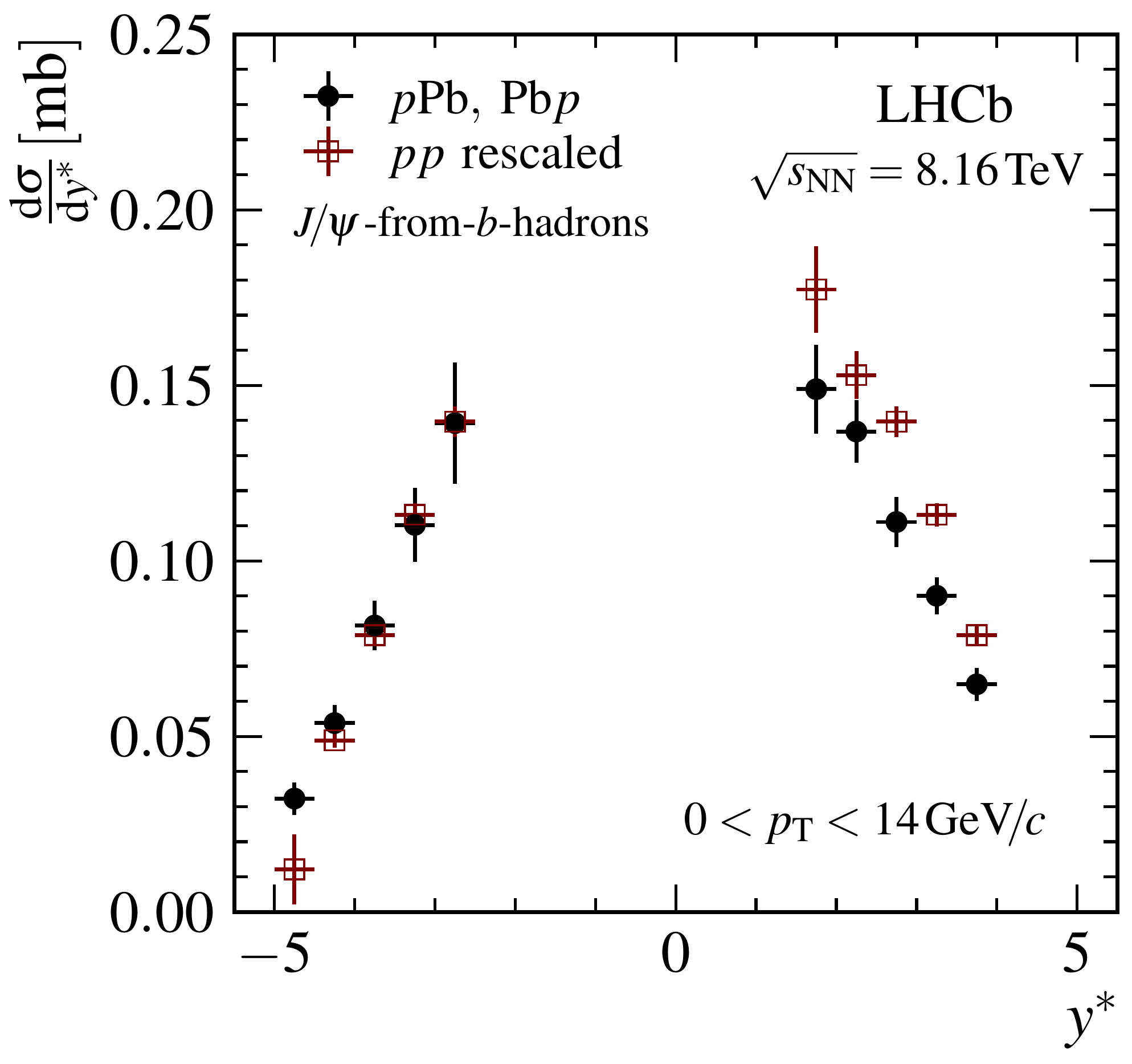}
\end{minipage}
\caption{Absolute production cross-sections of (left) prompt $J/\psi$ and (right) $J/\psi$-from-$b$-hadrons, as a function of $y^{*}$, integrated
over the range $0 < p_{\rm T} <14 GeV/c$.}
\label{figure32}
\end{figure}

\begin{figure}[!htb]
\centering
\begin{minipage}[t]{0.49\textwidth}
\centering
\includegraphics[height=2in]{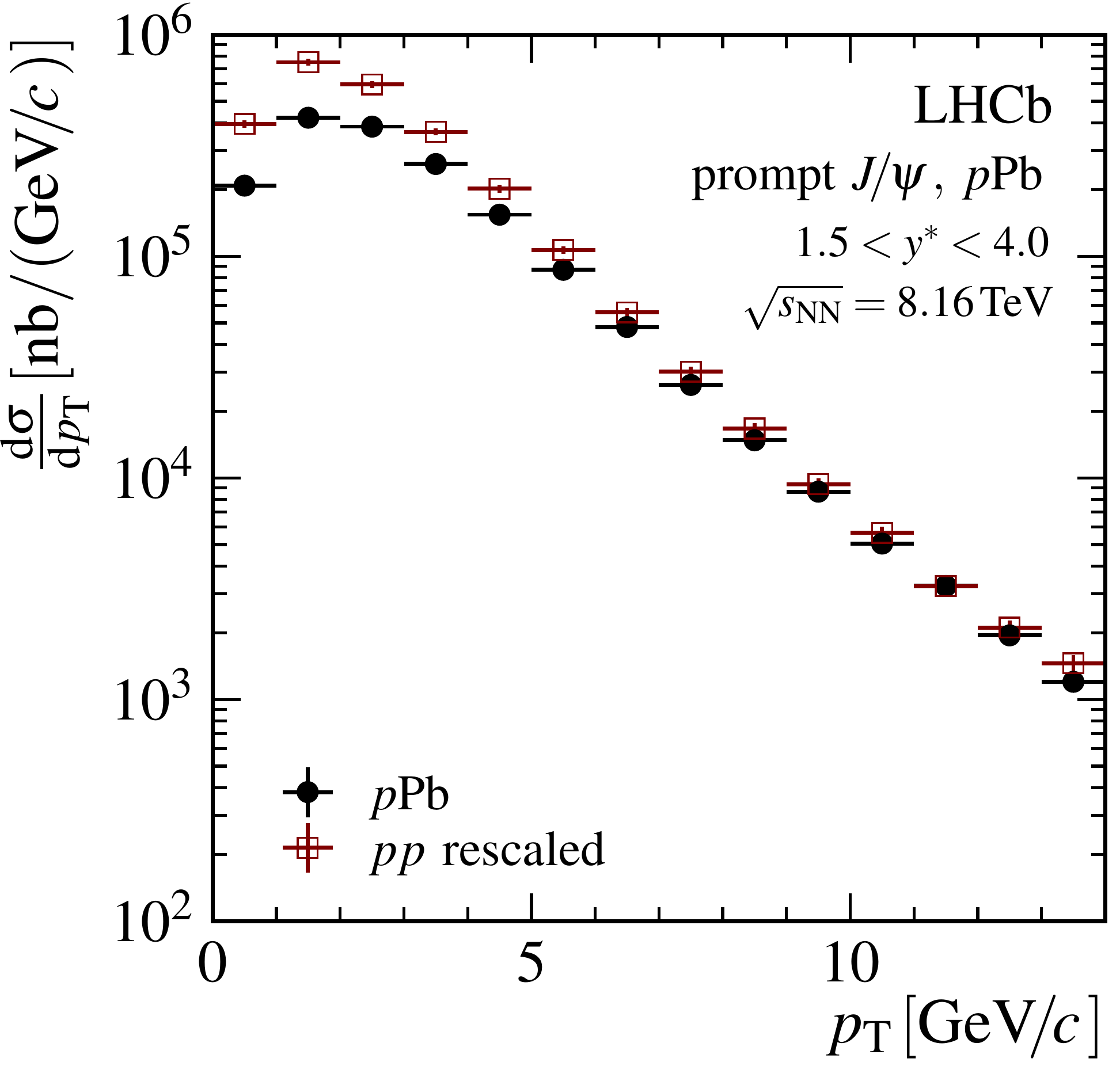}
\end{minipage}
\begin{minipage}[t]{0.49\textwidth}
\centering
\includegraphics[height=2in]{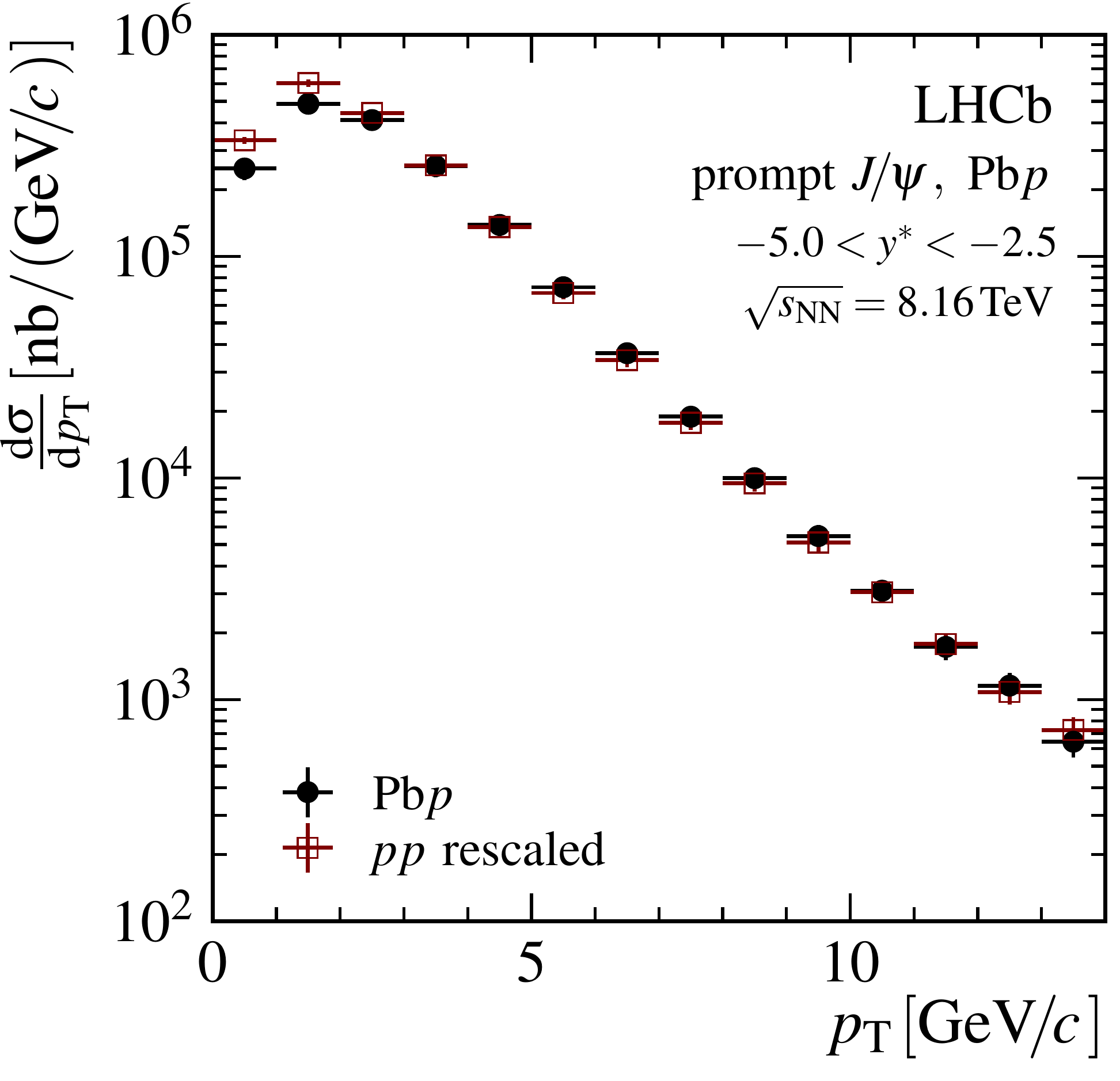}
\end{minipage}
\centering
\begin{minipage}[t]{0.49\textwidth}
\centering
\includegraphics[height=2in]{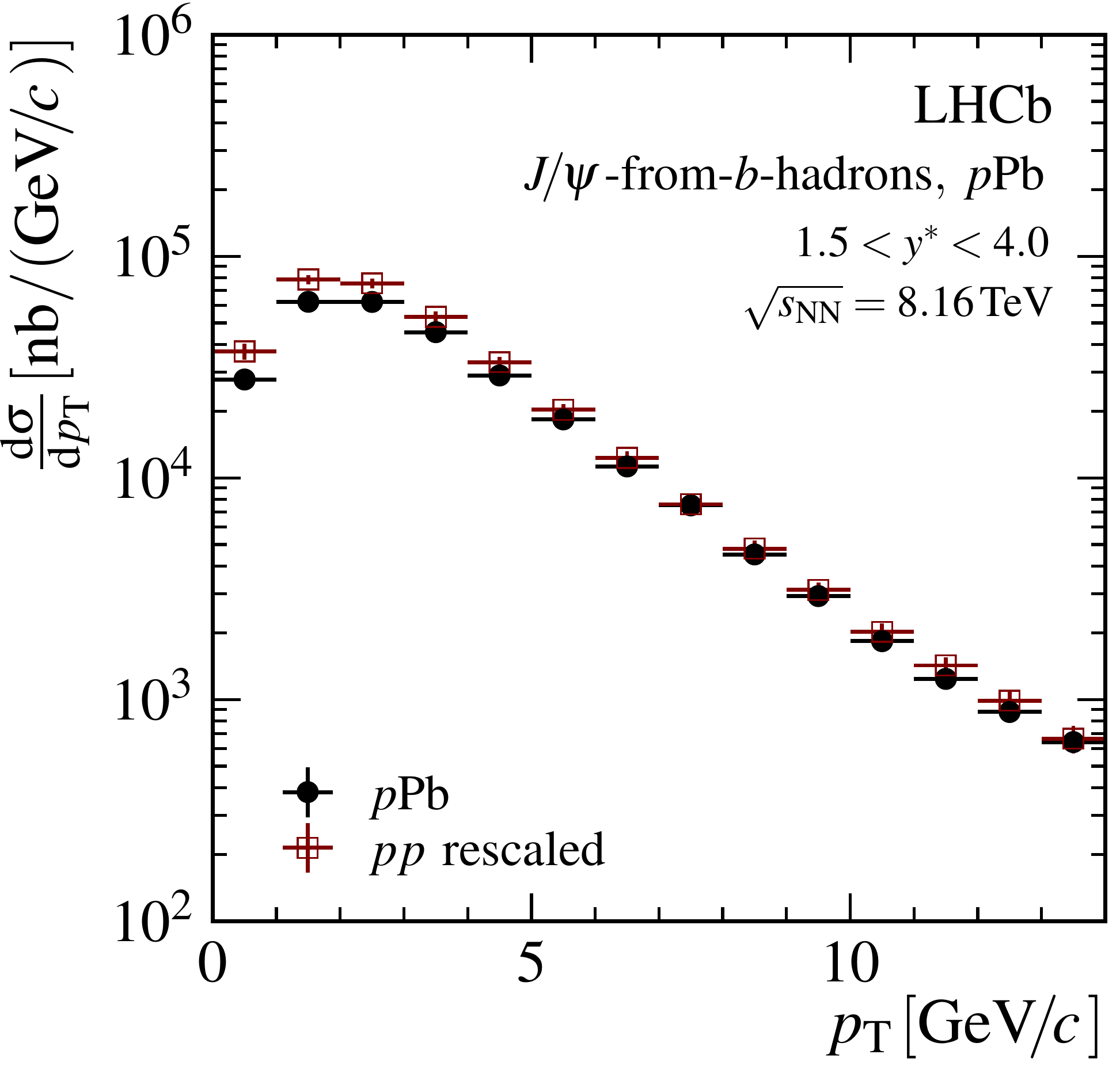}
\end{minipage}
\begin{minipage}[t]{0.49\textwidth}
\centering
\includegraphics[height=2in]{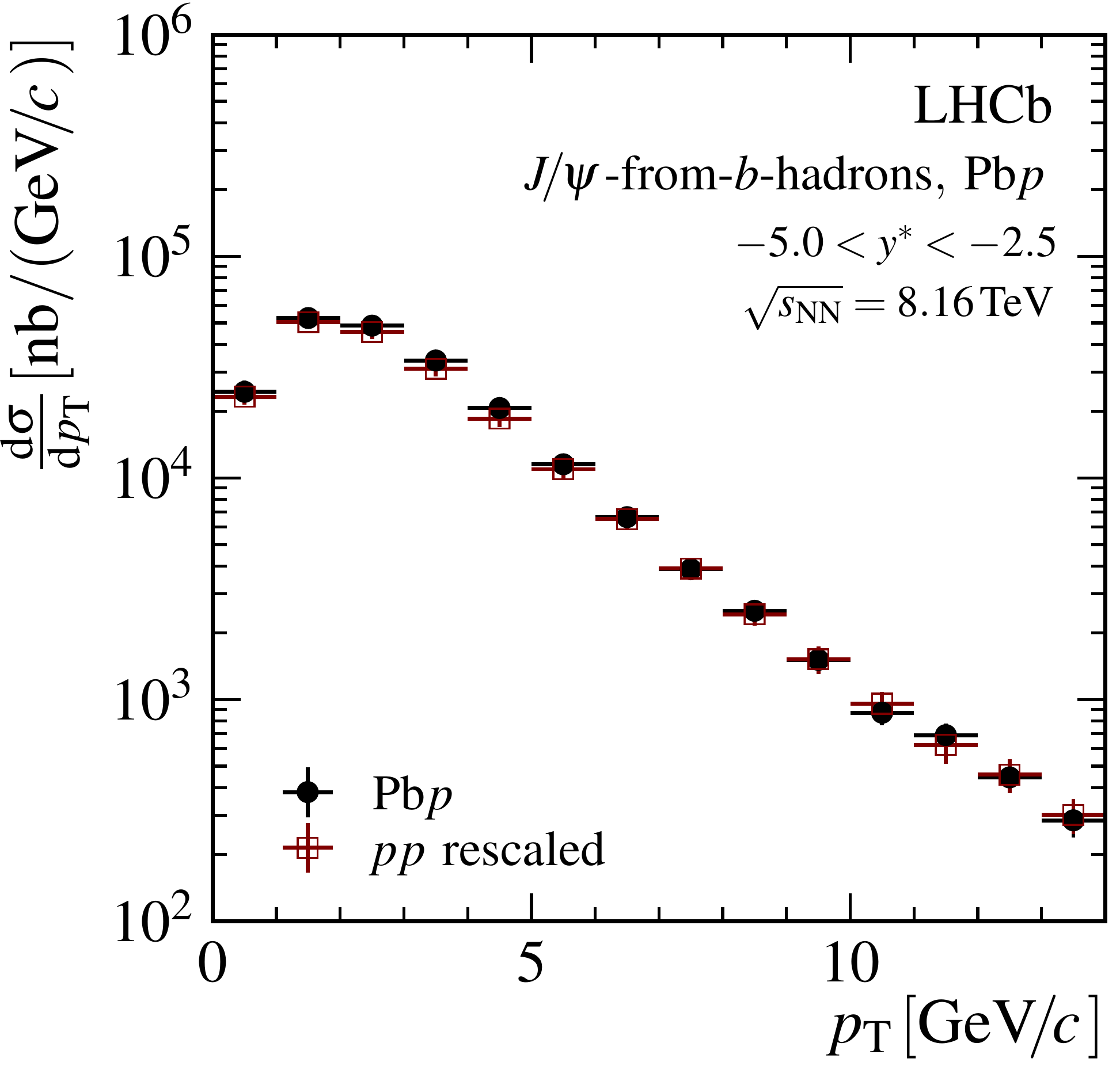}
\end{minipage}
\caption{Absolute production cross-sections of (top left) prompt $J/\psi$ in $p$Pb, (top right) prompt $J/\psi$ in Pb$p$,
(bottom left) $J/\psi$-from-$b$-hadrons in Pb$p$ and (bottom left) $J/\psi$-from-$b$-hadrons in Pb$p$, as a function of $p_{\rm T}$ and integrated over the rapidity range of the analysis.}	
\label{figure33}
\end{figure}

The single-differential cross-sections are compared with the ones from the inter/extrapolated $pp$ reference cross sections~\cite{8tevjpsi}. The results are shown in Fig.~\ref{figure32} and Fig.~\ref{figure33}.
There are some differences especially in the forward region and the low $p_{\rm T}$ region, which highlight
the effect of the nuclear environment.

\subsection{Nuclear modification factors}
\begin{figure}[!htb]
\centering
\begin{minipage}[t]{0.49\textwidth}
\centering
\includegraphics[height=2in]{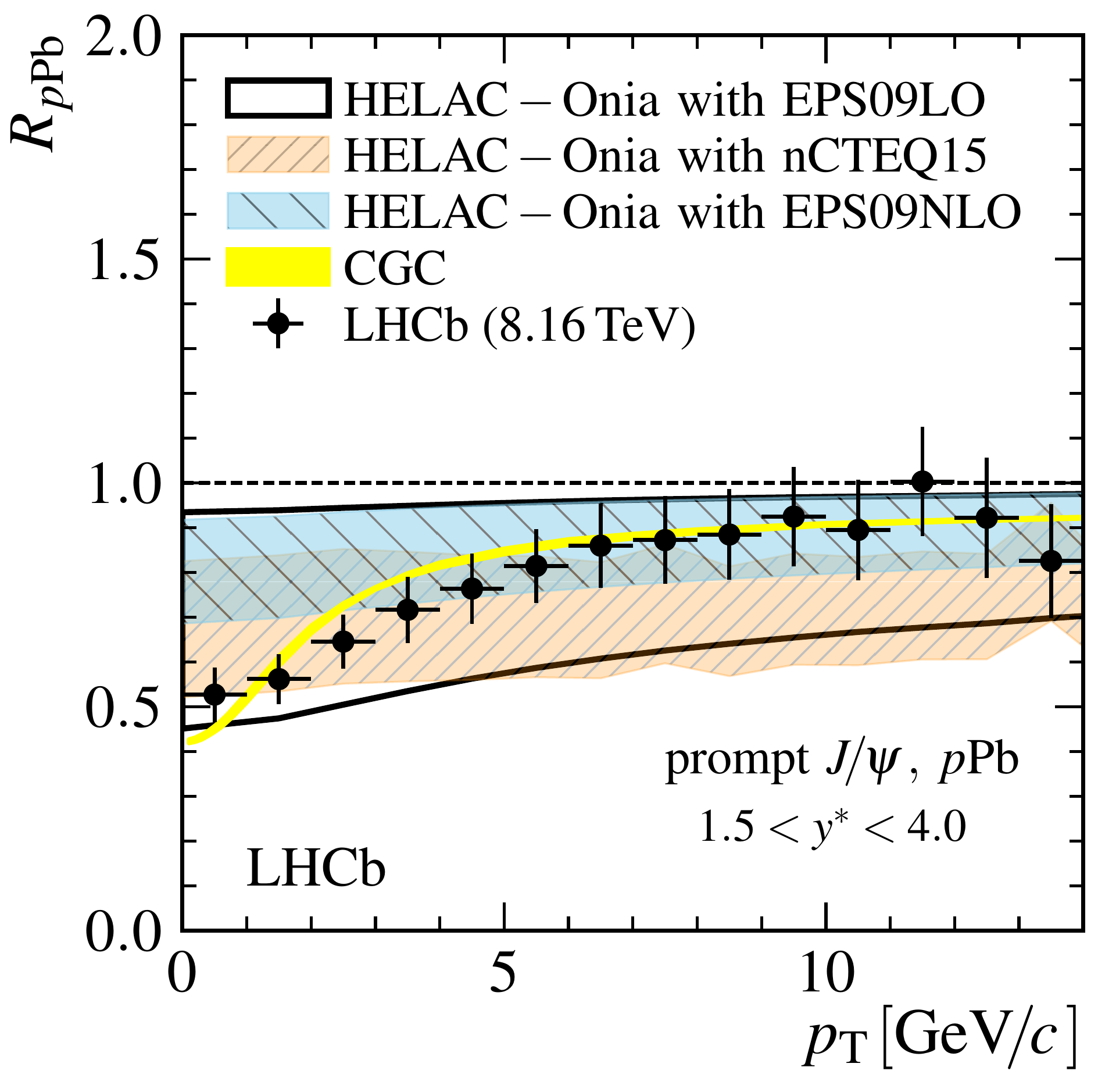}
\end{minipage}
\begin{minipage}[t]{0.49\textwidth}
\centering
\includegraphics[height=2in]{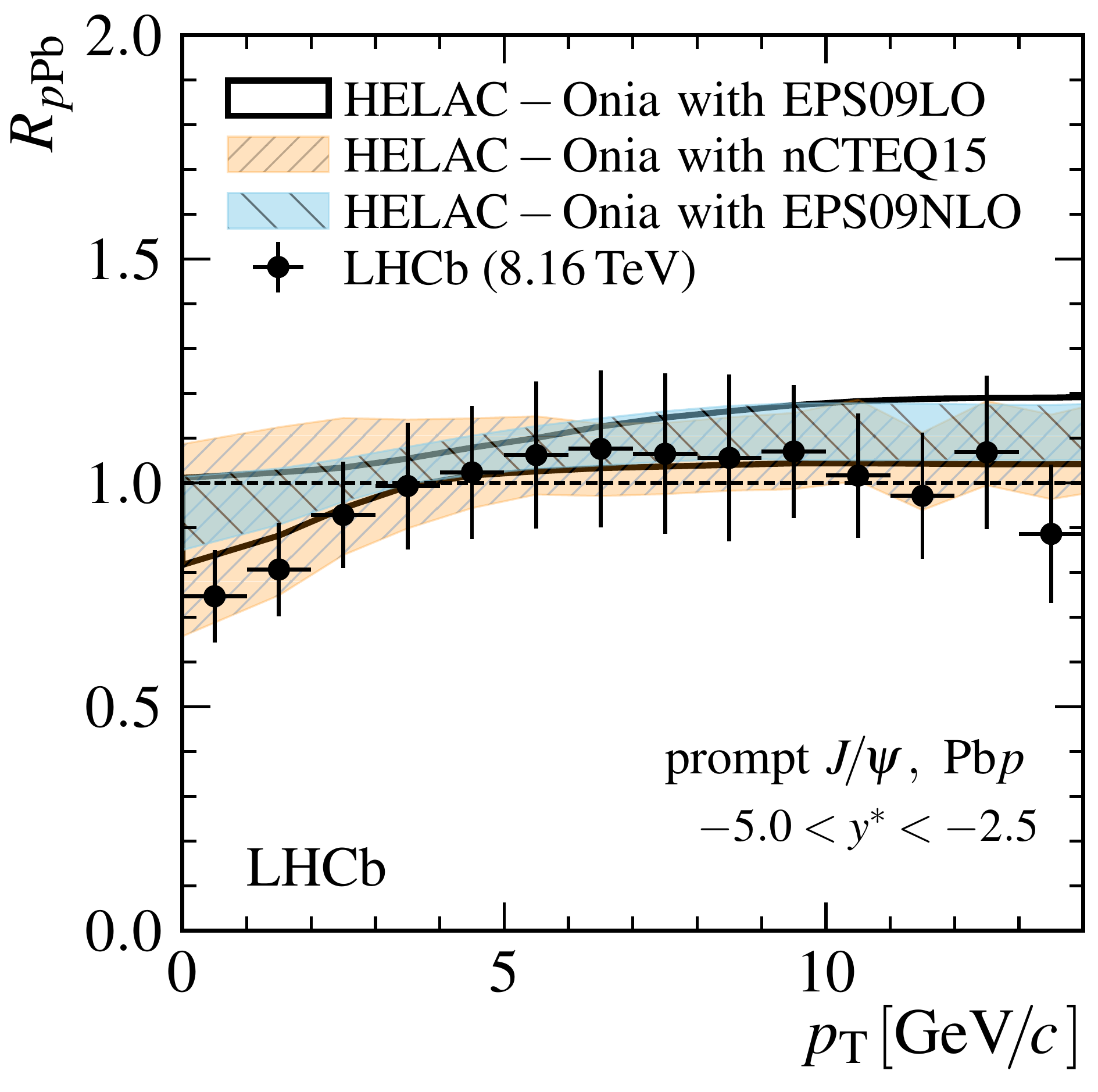}
\end{minipage}
\centering
\begin{minipage}[t]{0.49\textwidth}
\centering
\includegraphics[height=2in]{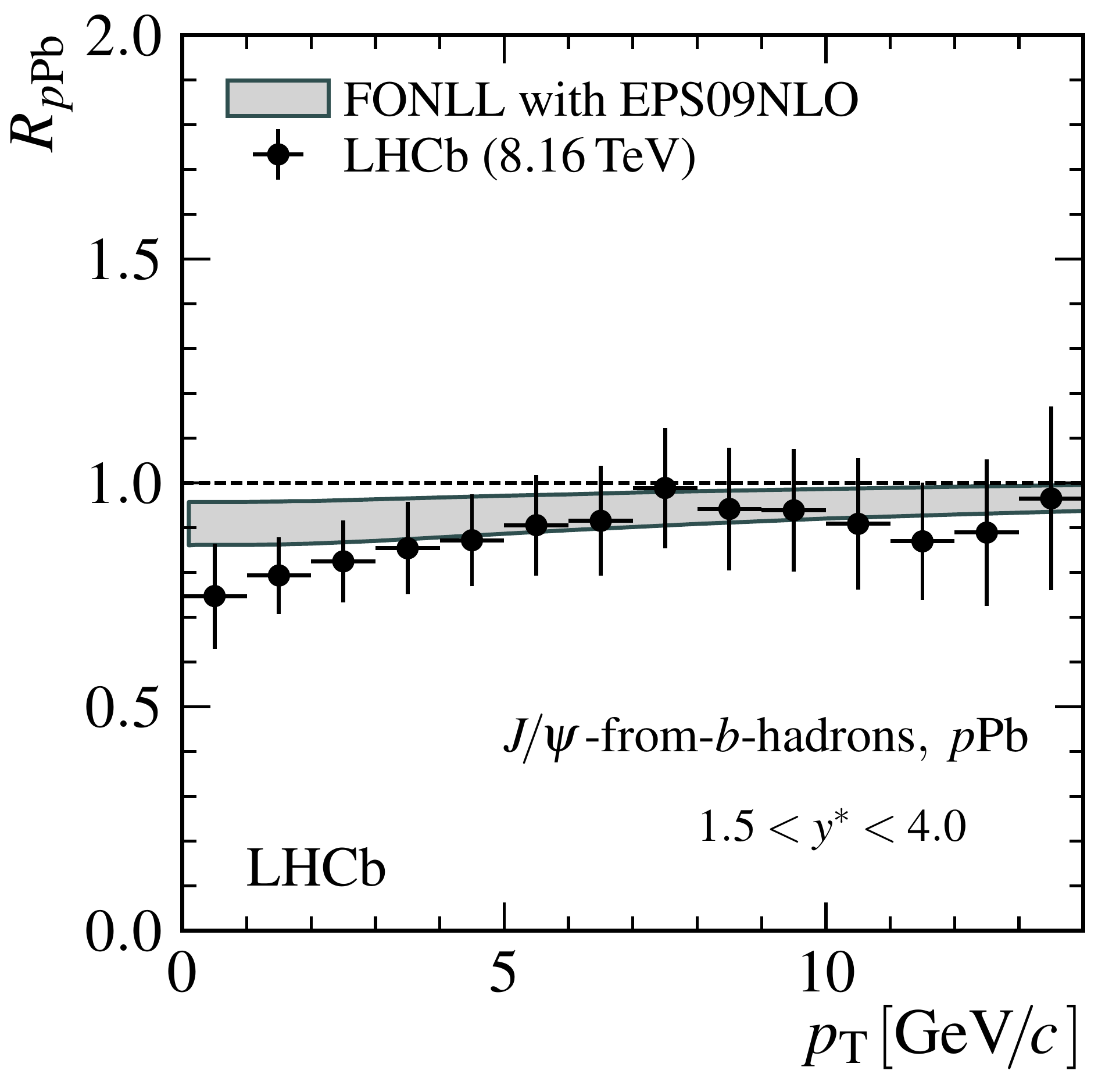}
\end{minipage}
\begin{minipage}[t]{0.49\textwidth}
\centering
\includegraphics[height=2in]{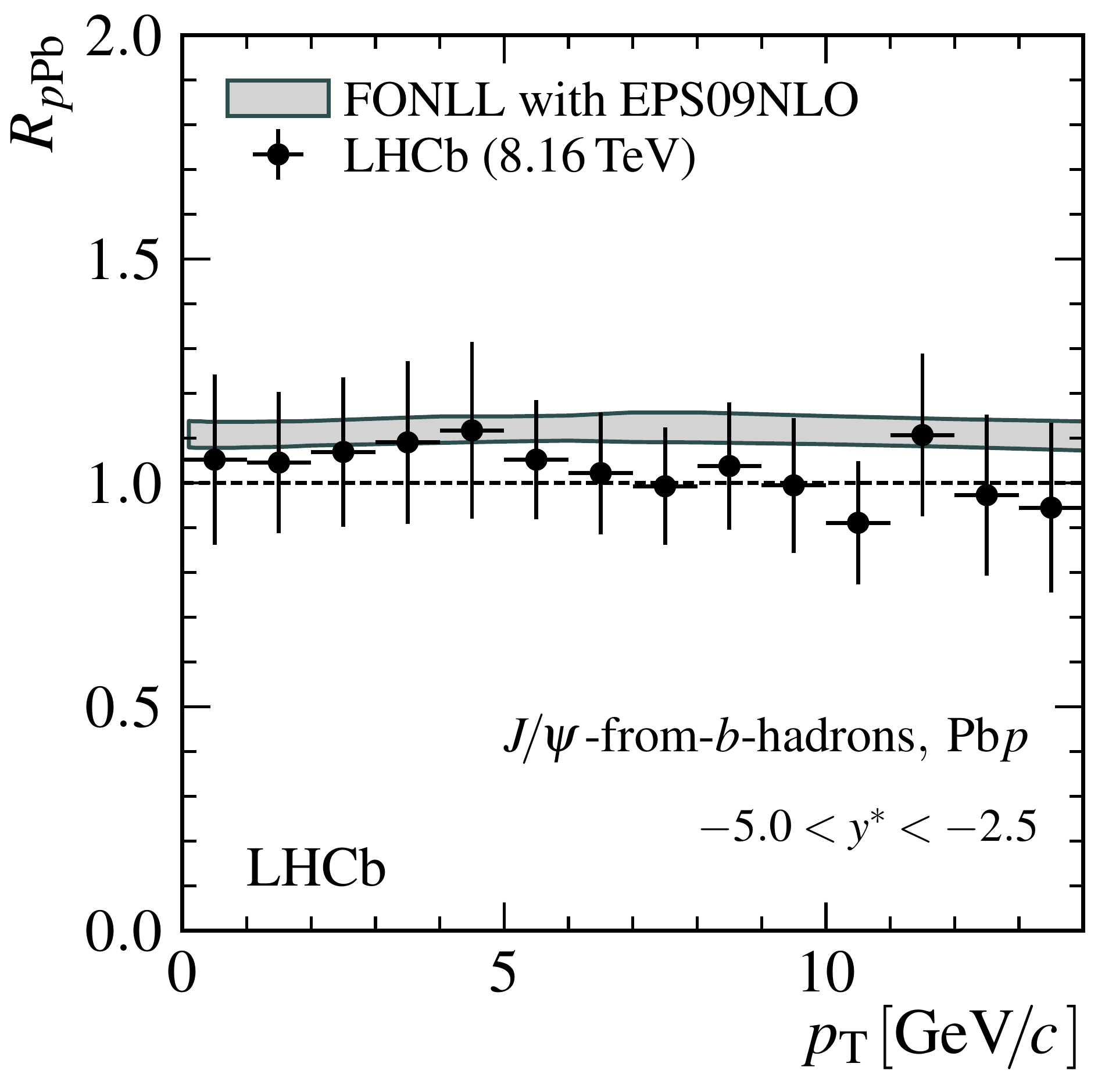}
\end{minipage}
\caption{$J/\psi$ nuclear modification factor, $R_{p{\rm Pb}}$, integrated over $y^{*}$ in the analysis     range,  
as a function of $p_{\rm T}$ for (top left) prompt $J/\psi$ in $p$Pb, (bottom left) $J/\psi$-from-$b$-hadrons in $p$Pb, (top right) prompt $J/\psi$ in Pb$p$
and (bottom right) $J/\psi$-from-$b$-hadrons in Pb$p$.}
\label{figure4}
\end{figure}

\begin{figure}[!htb]
\centering
\begin{minipage}[t]{0.49\textwidth}
\centering
\includegraphics[height=2in]{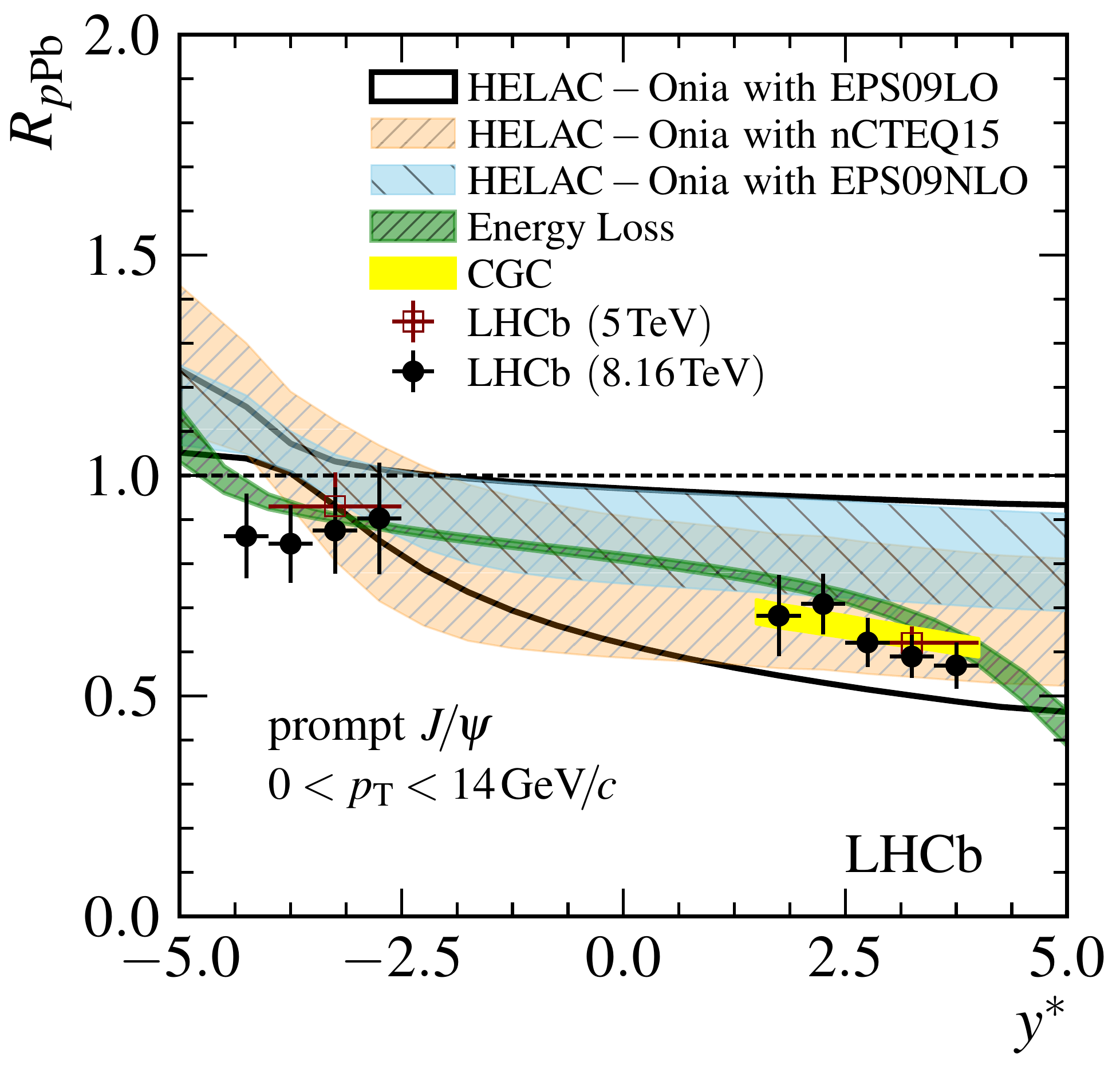}
\end{minipage}
\begin{minipage}[t]{0.49\textwidth}
\centering
\includegraphics[height=2in]{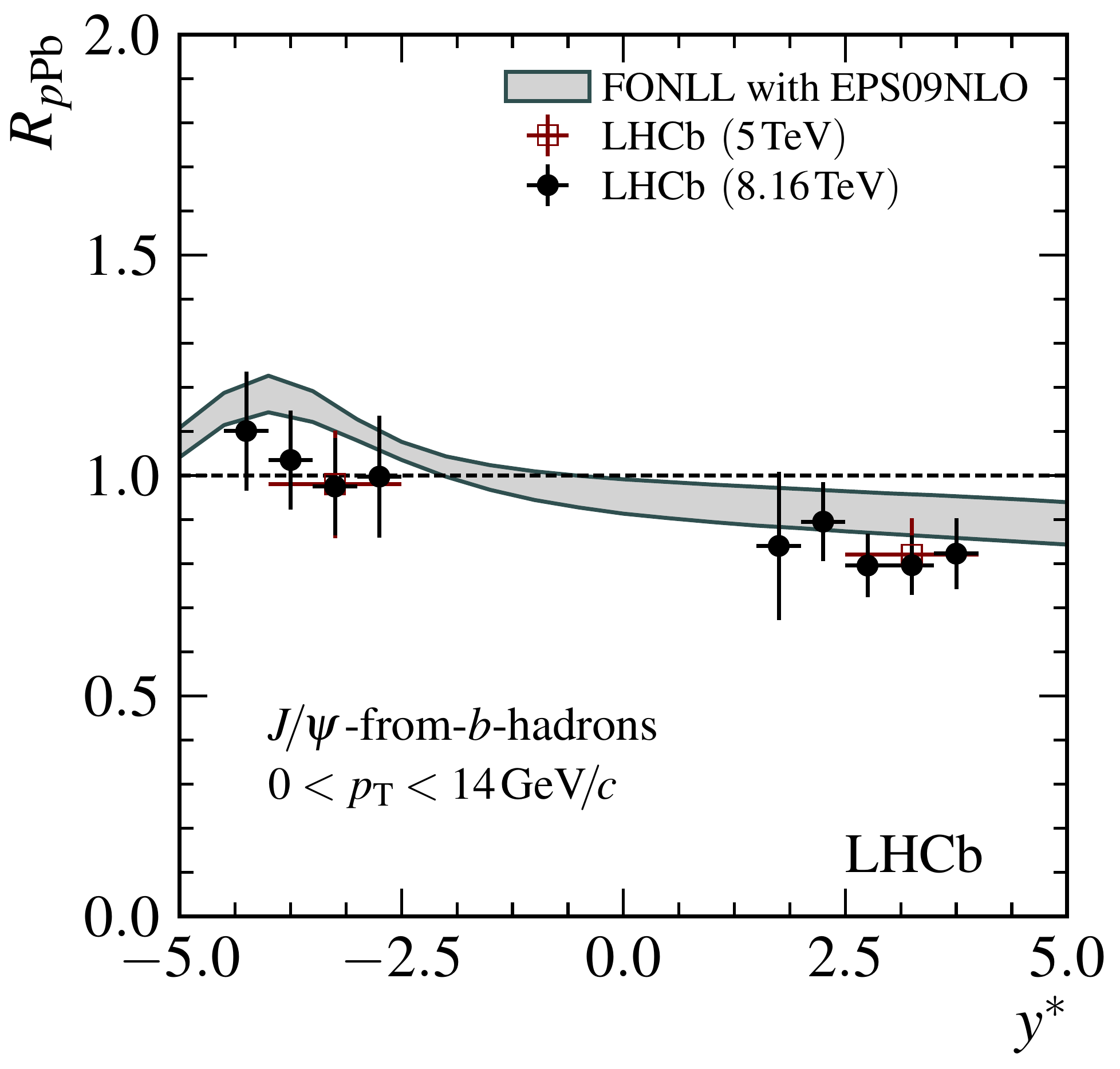}
\end{minipage}
\caption{$J/\psi$ nuclear modification factor, $R_{p{\rm Pb}}$, integrated over $p_{\rm T}$             
as a function of $y^{*}$ for (left) prompt $J/\psi$ and (right) $J/\psi$-from-$b$-hadrons.}	
\label{figure41}
\end{figure}

The nuclear modification factors for prompt $J/\psi$ and nonprompt $J/\psi$ are shown in Fig.~\ref{figure4} and Fig.~\ref{figure41}.

At forward rapidity, $1.5<y^{*}<4.0$, a strong suppression of up to $50\%$ is observed in the case of prompt $J/\psi$ at low transverse momentum.
With increasing $p_{\rm T}$, $R_{p\rm Pb}$ approaches unity and the suppression is stronger
at more forward rapidities.
The production of $J/\psi$-from-b-hadrons is also suppressed
compared to that in $pp$ collisions.
No dependence as a function of rapidity is observed within the
experimental uncertainties. The dependence as a function of the transverse momentum is
weaker for $J/\psi$-from-b-hadrons compared to prompt $J/\psi$.

At backward rapidity, $-5.0<y^{*}<-2.5$, a weaker suppression of prompt $J/\psi$ 
production at low transverse momentum is observed, of up to $25\%$. Similarly to the
forward rapidity region, the suppression is weakening and the nuclear modification factor is
approaching values consistent with unity at high transverse momentum. 

For the prompt $J/\psi$ results,
they are compared with calculations of different groups: collinear factorisation using different nPDFs
~\cite{Shao2, Shao3}, CGC effective field theory~\cite{jpsicgc, jpsicgc1} and coherent energy loss~\cite{loss}.
The CGC predictions agree with the behaviour of the data well
at forward rapidity and this is the same conclusion in the prompt $D^{0}$ measurement.
The HELAC-Onia calculations have been tuned to reproduce prompt $J/\psi$
cross-section measurements in $pp$ collision and then combined with different sets of nPDFs.
The large uncertainties caused by the missing experimental constraints on the gluon
density in the nucleus at low x.
At backward rapidities, the experimental points exhibit a different rapidity shape to the calculations and it's different with the shape of the prompt $D^{0}$ shown in Fig.~\ref{figure2} as well. 
The coherent energy loss model is able to provide the overall
shape of the suppression, but overestimates the experimental data at forward rapidities.

For the $J/\psi$-from-b-hadrons results, 
a perturbative QCD calculation at fixed-order next-to-leading
logarithms(FONLL)~\cite{fonll, fonll1} coupled with the EPS09 nPDF set at next-to-leading order
are compared with the data.
The measurements overall agree with the calculation within uncertainties.
However, the calculation tends to show larger nuclear
modification factors than the data and the
slope of the theoretical curve is not seen in the experimental data at backward rapidity.

Furthermore, the measurements are compatible with the 5 TeV results~\cite{5tevjpsi}.

\subsection{Forward-Backward production ratio}
\begin{figure}[!htb]
\centering
\begin{minipage}[t]{0.49\textwidth}
\centering
\includegraphics[height=2in]{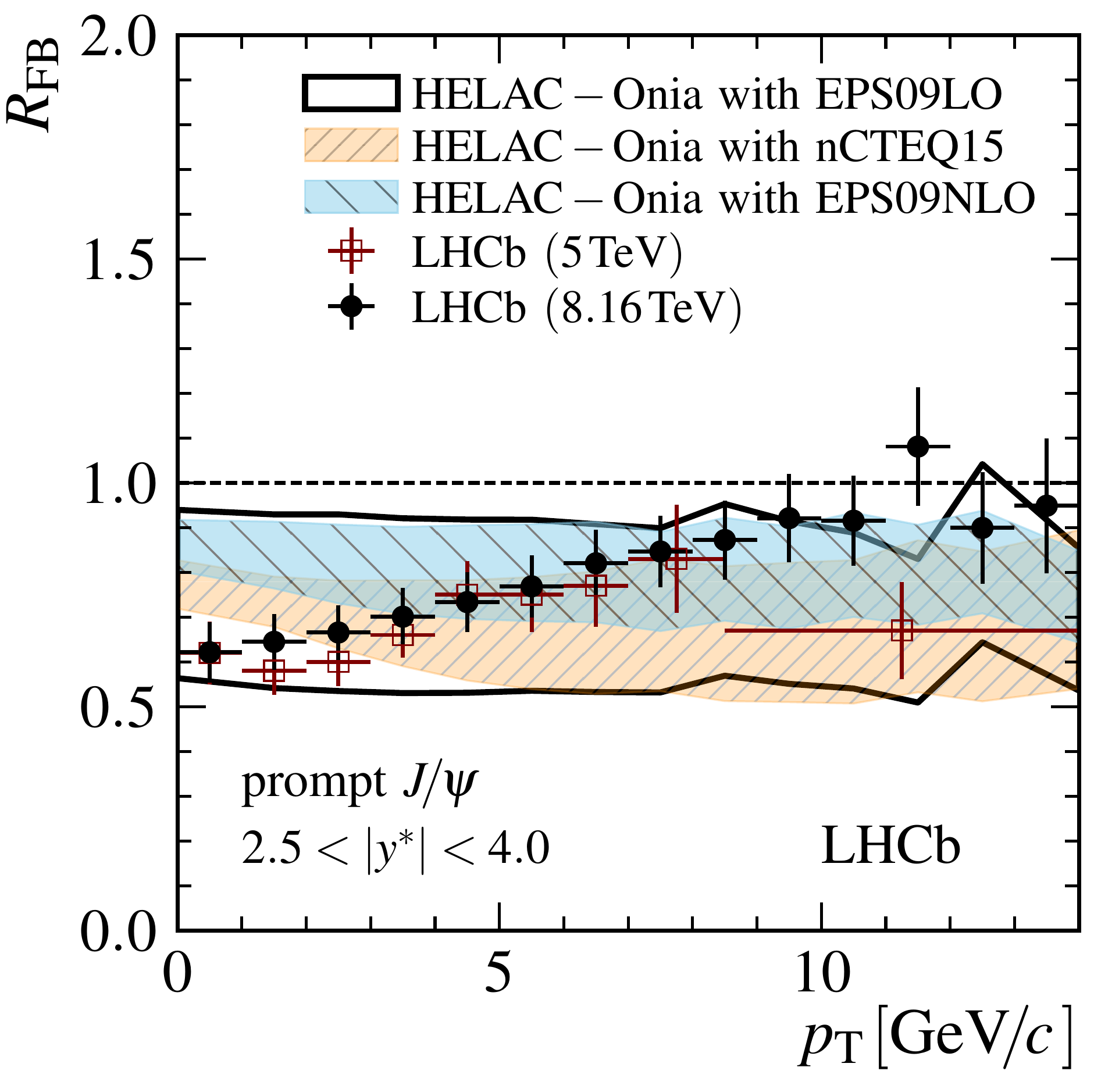}
\end{minipage}
\begin{minipage}[t]{0.49\textwidth}
\centering
\includegraphics[height=2in]{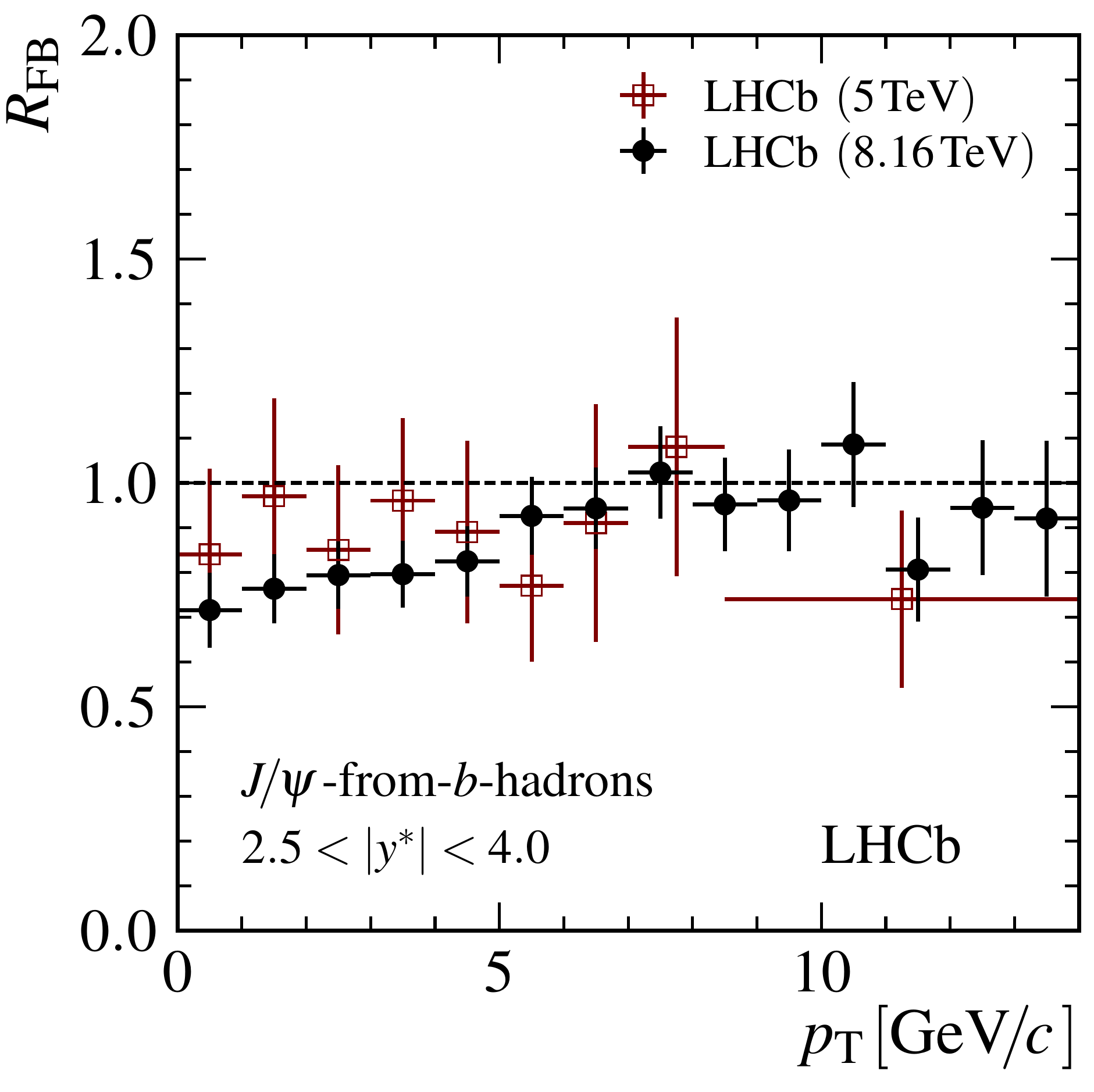}
\end{minipage}
\caption{Forward-to-backward ratios, $R_{{\rm FB}}$, integrated over the common rapidity range $2.5 <|y^{*}|< 4.0$
as a function of $p_{\rm T}$ for (left) prompt $J/\psi$ and (right) $J/\psi$-from-$b$-hadrons.}
\label{figure50}
\end{figure}
\begin{figure}[!htb]
\centering
\begin{minipage}[t]{0.49\textwidth}
\centering
\includegraphics[height=2in]{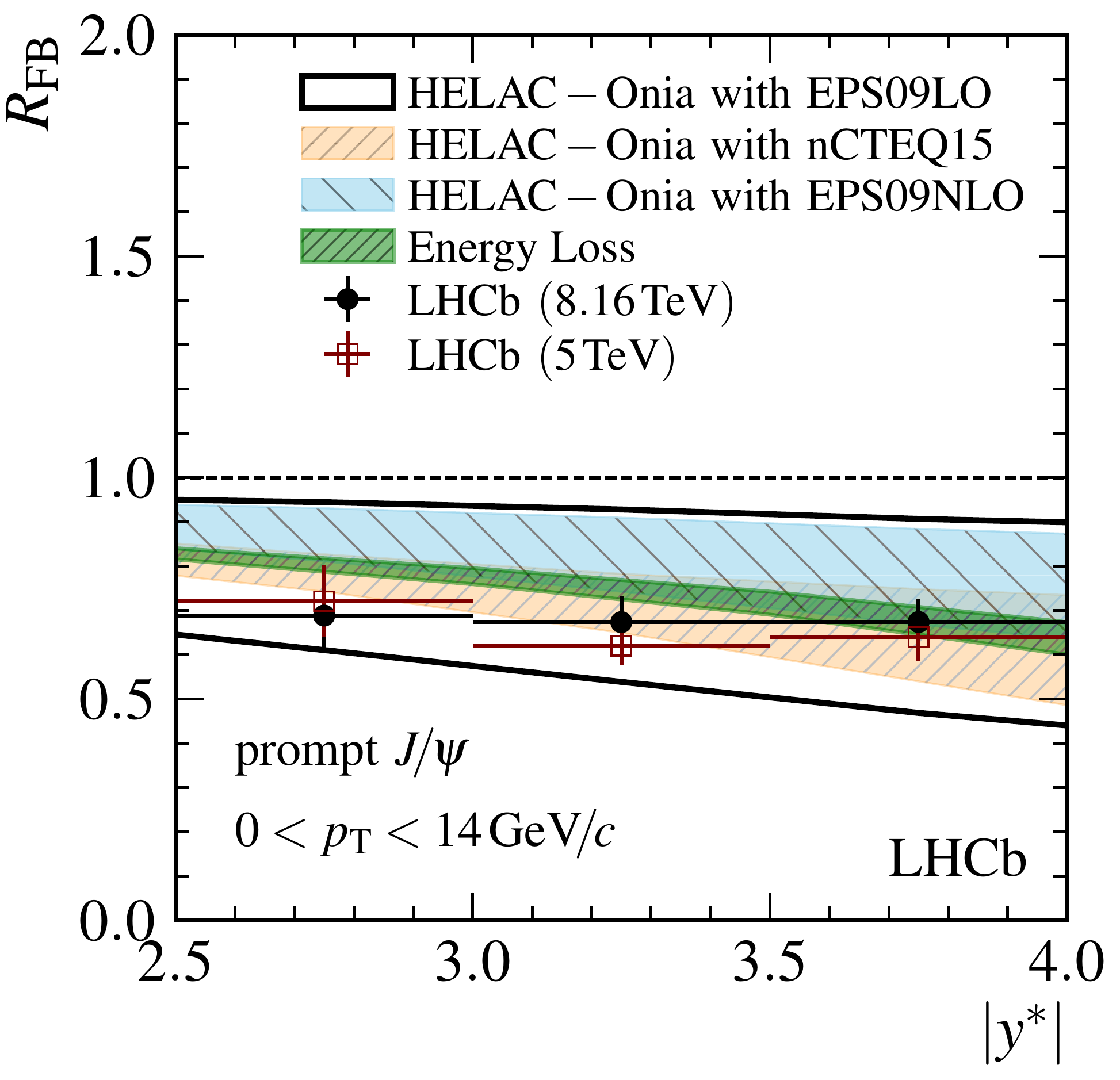}
\end{minipage}
\begin{minipage}[t]{0.49\textwidth}
\centering
\includegraphics[height=2in]{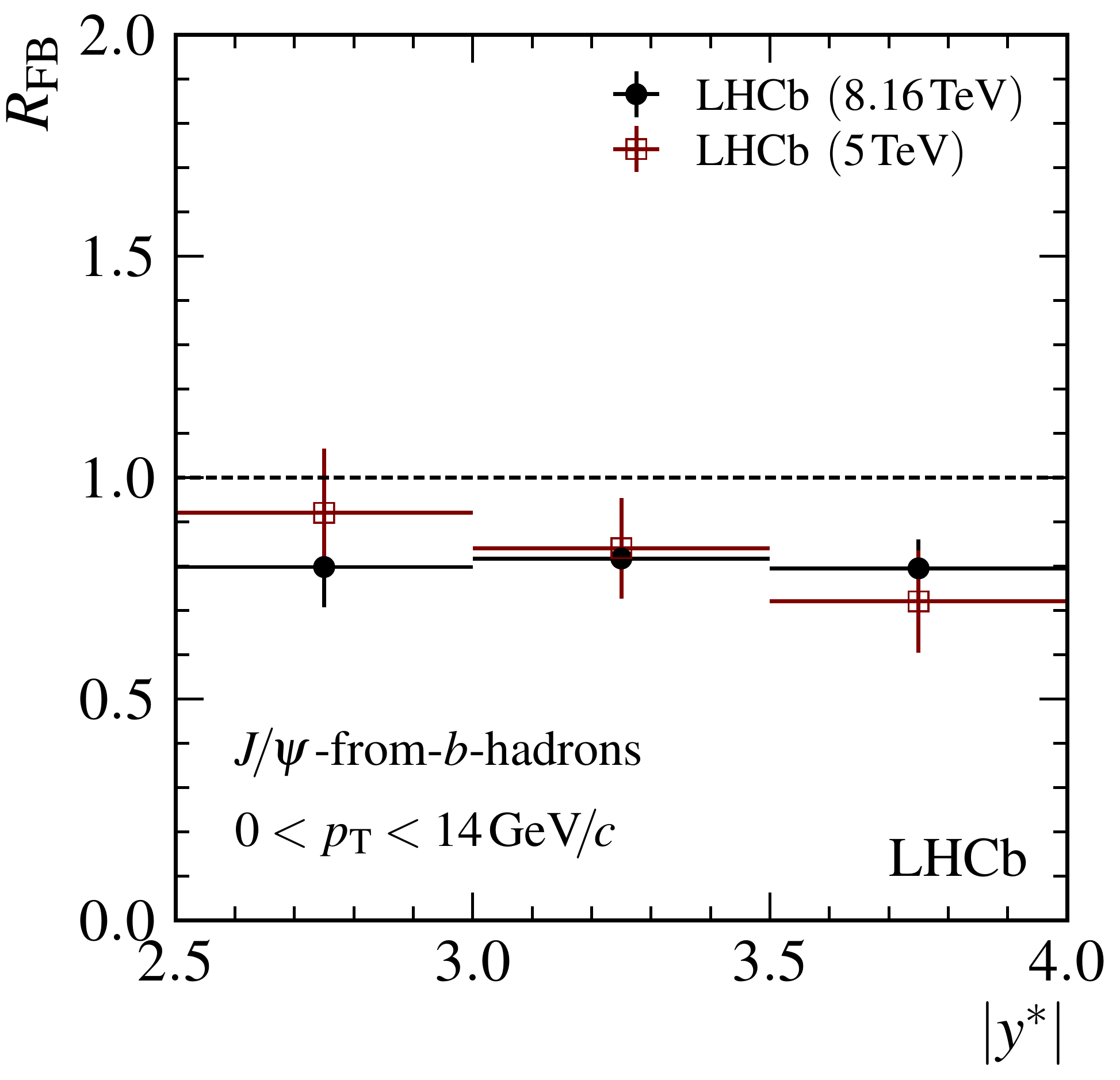}
\end{minipage}
\caption{Forward-to-backward ratios, $R_{{\rm FB}}$, integrated over $p_{\rm T}$ in the range $0<p_{\rm T} <14$ GeV/c
as a function of $|y^{*}|$ for (left) prompt $J/\psi$ and (right) $J/\psi$-from-$b$-hadrons.}	
\label{figure51}
\end{figure}

For prompt $J/\psi$ and nonprompt $J/\psi$, we also measure the forward to backward production ratio 
$R_{\rm FB}(p_{\rm T},y^{*})$. 
The measurements as a function of $p_{\rm T}$ or $y^{*}$ are shown in Fig.~\ref{figure50} and Fig.~\ref{figure51} in comparison with those in $p$Pb collisions at $\sqrt{s}$=5 TeV~\cite{5tevjpsi} and theoretical calculations.

The measurements of $R_{\rm FB}(p_{\rm T},y^{*})$ at $\sqrt{s}$=5 TeV and at $\sqrt{s}$=8.16 TeV
are found to be in agreement.
The experimental results agree with the predictions based on HELAC-onia with different nPDFs within uncertainties for prompt $J/\psi$.
The coherent energy loss calculation is compared with
the rapidity dependence of the experimental data points. The slope is 
slightly differs from the one observed with
the experimental data points.

\section{Summary}
LHCb has collected large data samples in proton-lead collisions.
A new measurement of prompt $D^{0}$ production in $p$Pb and Pb$p$ collisions at $\sqrt{s}$=5 TeV shows strong cold nuclear matter effects that  
can be described by 
calculations using nPDFs, CGC framework.
The first result for $J/\psi$ production in $p$Pb and Pb$p$ collisions at $\sqrt{s}$= 8.16 TeV at LHC has been measured. Prompt and non-prompt $J/\psi$ production 
results show 
clear suppression and confirm the findings at $\sqrt{s}$=5 TeV~\cite{5tevjpsi}
with higher precision.
Theoretical calculations for the nuclear modification
factor based on collinear factorisation with different nuclear parton distribution functions,
coherent energy loss as well as 
the color glass condensate approach calculation
can account for the
majority of the observed dependencies for prompt $J/\psi$.
Calculations from the FONLL coupled with the EPS09 nPDF set can overall describe the data points within uncertainties for the non-prompt $J/\psi$.


\begin{thebibliography}{99}

\bibitem{5tevD} 
  R. Aaij {\it et al.}  [LHCb Collaboration],
  Study of prompt $D^{0}$ meson production in $p$Pb collisions at $\sqrt{s}=5$ TeV,
  Accepted by JHEP,
  [arXiv:1707.02750].
 
\bibitem{8.16tevjpsi} 
  R. Aaij {\it et al.}  [LHCb Collaboration],
	Prompt and nonprompt $J/\psi$ production and nuclear modification in $p$Pb collisions at $\sqrt{s}=8.16$ TeV,
  Phys.\ Lett.\ B {\bf 774} (2017) 159-178,
	[arXiv:1706.07122].

\bibitem{nPDFs} 
  H.Fujii, F. Gelis, and R. Venugopalan,
  Quark pair production in high energy pA collisions: General features,
  Nucl.\ Phys.\ A {\bf 780}, 146 (2006)
  [arXiv:0603099 [hep-ph]].
 
\bibitem{CGC} 
  D. Kharzeev and K. Tuchin, Signatures of the color glass condensate in $J/\psi$ production off nuclear targets,
  Nucl.\ Phys.\ A {\bf 770}, 40 (2006)
  [arXiv:0510358 [hep-ph]].
 
\bibitem{loss} 
  F. Arleo and S.Peigne, Heavy-quarkonium suppression in pA collisions from parton energy loss in cold QCD matter,
  JHEP \ {\bf 03}, (2013) 122
  [arXiv:1212.0434].

 \bibitem{lhcb} 
  A. A. Alves Jr. {\it et al.}  [LHCb Collaboration],
  The LHCb detector at the LHC,
	JINST 3 (2012) S08005.
 
\bibitem{lhcb1} 
  R. Aaij {\it et al.}  [LHCb Collaboration],
  LHCb detector performance,
  Int.\ J.\ Mod.\ Phys.\ A {\bf 30} (2015) 1530022,
  [arXiv:1412.6352].

 \bibitem{eps09} 
  K. J. Eskola, H. Paukkunen, and C. A. Salgado,
  EPS09: A New generation of NLO and LO Nuclear Parton Distribution Functions,
	JHEP {\bf 04}, (2009) 065,
  [arXiv:0902.4154].
 
 \bibitem{cteq} 
  K. Kovarik {\it et al.},
  nCTEQ15 - Global analysis of nuclear parton distributions with uncertainties in the CTEQ framework,
	Phys.\ Rev.\ D {\bf 93} (2016), 085037,
  [arXiv:1509.00792].

\bibitem{Shao1} 
 J.-P. Lansberg and H.-S. Shao,
  Towards an automated tool to evaluate the impact of the nuclear modification of the gluon density on quarkonium, D and B meson production in proton-nucleus collisions,
	Eur. \ Phys.\ J.\ C {\bf 77} (2017), 1,
  [arXiv:1610.05382].
  
\bibitem{Shao2} 
  H.-S. Shao,
  HELAC-Onia 2.0: an upgraded matrix-element and event genera-tor for heavy quarkonium physics,
	Comput.\ Phys.\ Commun. {\bf 198} (2016) 238,
  [arXiv:1507.03435].
  
\bibitem{Shao3} 
  H.-S. Shao,
  HELAC-Onia: An automatic matrix element generator for heavy quarkonium physics,
  Comput.\ Phys.\ Commun. {\bf 184} (2013) 2562,
	[arXiv:1212.5293].
  
 \bibitem{fonll} 
  M. Cacciari, M. Greco, and P. Nason,
  The $p_{\rm T}$ spectrum in heavy-flavour hadroproduction,
	JHEP {\bf 05} (1998) 007,
	[arXiv:hep-ph/9803400].
 
 \bibitem{fonll1} 
  M. Cacciari, S. Frixione, and P. Nason,
  The $p_{\rm T}$ spectrum in heavy-flavour hadroproduction,
	JHEP {\bf 03} (2001) 006,
	[arXiv:hep-ph/0102134].
   
 \bibitem{jpsicgc} 
  B. Duclou\'e, T. Lappi, and H. M\"antysaari,
  Forward $J/\psi$ production in proton-nucleus collisions at high energy,
	Phys.\ Rev.\ D {\bf 91} (2015) 114005,
  [arXiv:1503.02789].
 
\bibitem{jpsicgc1} 
  B. Duclou\'e, T. Lappi, and H. M\"antysaari,
  Forward $J/\psi$ production at high energy: centrality dependence and mean transverse momentum,
	Phys.\ Rev.\ D {\bf 94} (2016) 074031,
  [arXiv:1605.05680].

\bibitem{8tevjpsi} 
  R. Aaij {\it et al.}  [LHCb Collaboration],
  Production of $J/\psi$ and $\Upsilon$ mesons in $pp$ collisions at $\sqrt{s}=8$ $TeV$, 
	JHEP {\bf 06} (2013) 064,
  [arXiv:1304.6977].

\bibitem{5tevjpsi}
 R. Aaij {\it et al.}  [LHCb Collaboration],
 Study of $J/\psi$ production and cold nuclear matter effects in $p$Pb collisions at $\sqrt{s}=5$ $TeV$,
 JHEP {\bf 02} (2014) 072,
 [arXiv:1308.6729].
\end{thebibliography}
\end{document}